\documentclass[prd, preprint, tightenlines, a4paper, eqsecnum, nofootinbib]{revtex4-2}
\usepackage[top=2cm, bottom=2cm, outer=2.5cm, inner=2.5cm]{geometry}
\usepackage[T1]{fontenc}

\usepackage{braket}
\usepackage{mathtools}
\usepackage{amssymb}
\usepackage{amsthm}
\usepackage{amsmath}
\usepackage{mathrsfs}
\usepackage{stackrel}
\usepackage{mdframed}
\usepackage{tabularx}

\usepackage{graphicx}
\usepackage{float}

\usepackage{url}
\usepackage{hyperref}

\DeclareMathOperator{\Imag}{Im}
\DeclareMathOperator{\Real}{Re}
\newcommand{\mini}{\scriptscriptstyle}
\makeatother

\begin{document}


\title{Ground and thermal states for the Klein-Gordon field  on a massless hyperbolic black hole with applications to the anti-Hawking effect}

\author{Lissa de Souza Campos}
\email{lissa.desouzacampos01@universitadipavia.it}
\author{Claudio Dappiaggi}
\email{claudio.dappiaggi@unipv.it}
\affiliation{Dipartimento di Fisica, Universit\`a degli Studi di Pavia, Via Bassi, 6, 27100 Pavia, Italy}
\affiliation{Istituto Nazionale di Fisica Nucleare -- Sezione di Pavia, Via Bassi, 6, 27100 Pavia, Italy}

\date{\today}
\vspace{1cm}
\begin{abstract}
	On an $n$-dimensional, massless, topological black hole with hyperbolic sections, we construct the two-point function both of a ground state and of a thermal state for a real, massive, free scalar field arbitrarily coupled to scalar curvature and endowed with Robin boundary conditions at conformal infinity. These states are used to compute the response of an Unruh-DeWitt detector coupled to them for an infinite proper time interval along static trajectories. As an application, we focus on the massless conformally coupled case and we show, numerically, that the anti-Hawking effect, which is manifest on the three-dimensional case, does not occur if we consider a four-dimensional massless hyperbolic black hole. On the one hand, we argue that this result is compatible with what happens in the three- and four dimensional Minkowski spacetime, while, on the other hand, we stress that it generalizes existing results concerning the anti-Hawking effect on black hole spacetimes.\\

\end{abstract}

\maketitle
\newpage

\section{Introduction}


 	We investigate the behavior of an Unruh-DeWitt detector following static trajectories and interacting, for an infinite proper time interval, either with a ground state or with a KMS state of a real, massive scalar field on a massless hyperbolic black hole. The detector is modeled as a two-level system such that the interaction with the underlying quantum state is codified by a monopole-type Hamiltonian operator. We are interested in the amplitudes of excitation and de-excitation of the detector within first-order perturbation theory. Such amplitudes are characterized by the transition rate, which is nothing but the Fourier transform of the pull-back along the detector trajectory of the underlying two-point function of the quantum field.

 	The main reason of our interest towards this quantity is that the transition amplitudes are strongly interconnected with the so-called anti-Hawking effect \cite{Henderson:2019uqo,Brenna:2015fga,Garay:2016cpf}. This is a phenomenon which can be understood as follows. Suppose that an Unruh-DeWitt detector thermalizes with a quantum field at a given temperature, see e.g. \cite{Hodgkinson:2011pc,Louko:2007mu, Louko:2006zv,Ng:2014kha}. In general, such temperature is proportional to the proper acceleration of the detector and, when its response rate decreases while the temperature increases, we refer to it as {\em anti-Hawking effect}, or as {\em anti-Unruh effect} if one works on a flat spacetime. Such phenomenon has attracted lately a lot of attention and it turns out that, on Minkowski spacetime, for a detector interacting with the vacuum state of a massless scalar field theory, the anti-Unruh effect is manifest in the amplitude of de-excitations in the three-dimensional case, but it does not occur at all in the four-dimensional Minkowski spacetime, see Section \ref{sec:The transition rate on Minkowski spacetime}. On a BTZ black hole, the anti-Hawking effect is instead manifest for a detector interacting either with the state constructed from restricting the global AdS ground state to Rindler trajectories \cite{Henderson:2019uqo}, or with the intrinsic ground state constructed from the BTZ metric \cite{Campos:2020twd}. In this work we show that, similarly to \cite{Campos:2020twd}, also on massless hyperbolic black holes, the anti-Hawking effect is manifest only in three dimensions and if one considers a ground state. On the contrary, it does not occur if one works with a thermal state, or for both states in four dimensions.

 To enter more into the details of our work, we stress that massless hyperbolic black holes are static $n$-dimensional solutions of Einstein equations in vacuum with negative cosmological constant that have constant, negative sectional curvature. They can be seen as higher-dimensional generalizations of a static BTZ black hole \cite{Mann:1997iz,Smith:1997wx,Birmingham:1998nr}. These solutions enjoy several notable properties such as being classically stable under scalar perturbations \cite{Birmingham:2007yv}. In addition they do possess a timelike conformal boundary at infinity. This is of particular relevance since we will be considering on top of these backgrounds a real, massive scalar field with an arbitrary coupling to scalar curvature. Since the underlying spacetime is not globally hyperbolic, the dynamics of matter fields is not specified only by initial data, but one needs also to assign suitable boundary conditions at conformal infinity. As first advocated in  \cite{Ishibashi:2004wx}, when considering a scalar field, a natural choice lies in the Robin boundary conditions, which have been recently thoroughly studied for an $n$-dimensional AdS spacetime in \cite{Dappiaggi:2016fwc,Dappiaggi:2017wvj,Dappiaggi:2018xvw} and for BTZ spacetime in \cite{Bussola:2017wki,Bussola:2018iqj}. Observe that one can consider more general boundary conditions, {\it e.g.} \cite{Dappiaggi:2018pju}, although we shall not discuss them further in this work.

For these reasons, we shall be considering a generic real, massive scalar field on a massless hyperbolic black hole with arbitrary Robin boundary conditions and, as first step we investigate how to construct the two-point function of a ground and of a thermal state. Similarly to AdS and to BTZ spacetimes, see \cite{Dappiaggi:2016fwc,Bussola:2017wki}, we restrict our attention to a subclass of boundary conditions of Robin type, characterized by the absence of bound states, which hinder the construction of a full-fledged ground state. In this scenario, by using a standard mode decomposition and tools from spectral analysis and from the theory of ordinary differential equations of Sturm-Liouville type, we are able to construct the sought ground and thermal states. Notice that we are not the first to study a scalar field on these backgrounds \cite{Wang:2001tk} and vacuum fluctuations \cite{Morley:2018lwn}, including also Robin boundary conditions \cite{Morley:2020zcd}, have been discussed in the literature though only in the massless, conformally coupled case. Hence our results expand these previous works for manifold reasons. On the one hand, our framework encompasses Robin boundary conditions for a massive field with arbitrary coupling to scalar curvature and on any spacetime dimension $n\geq3$. On the other hand, we construct the two-point function, not only for a Hartle-Hawking-like state, which we refer to as {\em thermal/KMS state}, but also for a Boulware-like state, which we refer to as {\em ground state}. In addition, as mentioned above, our construction allows to investigate numerically the response of an Unruh-DeWitt detector. Therefore we are able to generalize previous results \cite{Henderson:2019uqo,Campos:2020twd} concerning the anti-Hawking effect. In other words, we consider a detector following static trajectories of supercritical proper accelerations ($a>\frac{1}{L}$, $L$ being the AdS radius), and interacting for an infinite proper time interval with states associated to a massless conformally coupled scalar field. We show, numerically, that the anti-Hawking effect is not manifest if the dimension is four, and we expect it not to be manifest for dimensions higher than three. We stress that, although we focus our attention on the massless, conformally coupled case for the study of the anti-Hawking effect, the numerical analysis could be similarly reproduced also in the massive scenario. The same applies for higher-dimensional cases, at the price of a heavier computational effort. A preliminary numerical analysis indicates that, for higher values of mass, the anti-Hawking effect is present and more marked in all those cases when it occurs in the massless scenario.

\vskip .2cm

The paper is organized as follows. First, in Section \ref{sec: Massless hyperbolic black holes}, we describe the geometry of massless hyperbolic black holes. Secondly, in Section \ref{sec:The Klein-Gordon equation}, we study the solutions of the Klein-Gordon equation on such class of backgrounds. Then, in Section \ref{sec:Physically Sensible States}, we construct, using considerations from spectral theory applied to second order differential equations of Sturm-Liouville type, the two-point function both of a ground state and of a thermal state. In Section \ref{sec:An Unruh-DeWitt detector}, we briefly describe the Unruh-DeWitt detector framework and we write an explicit expression for the transition rate. Finally, in Section \ref{sec:Numerical analysis}, as an application of the framework established, we perform a numerical analysis for the massless conformally coupled, real scalar field and we study the dependence of the transition rate on the energy gap, on the boundary condition, on the spacetime dimension and on the local Hawking temperature.


\section{Massless hyperbolic black holes}
	\label{sec: Massless hyperbolic black holes}


In this section, we introduce the geometric data of interest for this paper. From now on, with $(\mathcal{M},g)$ we denote an $n$-dimensional, $n\geq 3$, static solution of vacuum Einstein's equations with negative cosmological constant $\Lambda$ such that $(\mathcal{M},g)$ is isometric to $\mathbb{R}\times I\times\Sigma_{n-2}$. Here, $I\subseteq\mathbb{R}$ while $\Sigma_{n-2}$ is a simply-connected, complete Riemannian manifold of constant, negative Gaussian curvature which we normalize to $-1$. In other words $\Sigma_{n-2}$ is an $(n-2)$-dimensional hyperbolic space. The line element reads
\begin{equation}
	\label{Eq: metric}
ds^2=-f(r)dt^2+\frac{dr^2}{f(r)}+r^2d\Sigma^2_{n-2},
\end{equation}
where $t\in\mathbb{R}$ and $$d\Sigma^2_{n-2}=d\theta^2+\sinh^2\theta d\mathbb{S}^2_{n-3}(\varphi_1,\dots,\varphi_{n-3})$$ is the line element of the $(n-2)$-dimensional hyperboloid with $\theta\in\mathbb{R}$, while $d\mathbb{S}^2_{n-3}$ is the line element of the unit $(n-3)$-dimensional sphere endowed with the standard angular coordinates $\varphi_1,\dots,\varphi_{n-3}$. The function $f(r)$ in Equation \eqref{Eq: metric} reads
\begin{equation}
	\label{Eq: f(r) function}
	f(r)=-1-\frac{2M}{r^{n-3}}+\frac{r^2}{L^2},
\end{equation}
where $M\geq 0$ can be interpreted as the black hole mass while $L=\sqrt{\frac{(n-1)(n-2)}{-2\Lambda}}$ is the AdS radius which we normalize, for convenience, to $1$. From Equation \eqref{Eq: f(r) function}, we can infer that the coordinate $r$ has support in the interval $(r_h,\infty)$, where $r_h$ is such that $f(r_h)=0$. As $r\to\infty$, $(\mathcal{M},g)$ has a conformal timelike boundary proper of any asymptotically AdS spacetime.

For $M>0$, Equation \eqref{Eq: metric} together with Equation \eqref{Eq: f(r) function} identify the $n$-dimensional hyperbolic Schwarzschild-AdS spacetime. If $M=0$, the underlying background is no longer singular, although a coordinate singularity occurs at $r=r_h=1$. Since for $M=0$, such as in the case $M>0$, the locus $r=r_h$ is a bifurcate horizon generated by the Killing vector $\partial_t$, this limiting scenario is referred to as {\em massless black holes}, {\it cf.} \cite{Birmingham:1998nr,Mann:1997iz,Smith:1997wx}.

Observe that we could drop the assumption of $\Sigma_{n-2}$ being simply connected considering instead homogeneous manifolds obtained as the quotient between the $(n-2)$-dimensional hyperbolic space by the subspace obtained from the transitive action of any discrete subgroup of the underlying isometry group. We decided to discard these cases for the sake of simplicity and clarity of the presentation. Yet, a reader should bear in mind that, barring minor modifications, all our results can be extended to these cases. Moreover, due to the possibility of performing different compactifications on $\Sigma_{n-2}$, these solutions are also known as {\em topological black holes}, see {\it e.g.} \cite{Vanzo:1997gw,Klemm:1998bb}. Since we focus on the simply connected case, which is the hyperbolic space case, we refer to them as {\em hyperbolic black holes}.

In this paper we work solely on massless hyperbolic black holes, indicating the underlying background as $(\mathcal{M},g_0)$ to highlight that the metric is the same as in Equation \eqref{Eq: metric} though considering $M=0$ in Equation \eqref{Eq: f(r) function}.


\section{The Klein-Gordon equation}
	\label{sec:The Klein-Gordon equation}


In this section, we consider a real, massive scalar field $\Psi:\mathcal{M}\to\mathbb{R}$, arbitrarily coupled to scalar curvature. The dynamics is ruled by the Klein-Gordon equation
	\begin{equation}
		\label{eq:KG equation}
		P\Psi :=\left(\Box_{g_0}- \mu^2\right)\Psi=0,
	\end{equation}
where $\Box_{g_0}$ is the D'Alembert wave operator built out of \eqref{Eq: metric}:
\begin{equation}
	\label{eq:dalembertian for the massless bh}
	\Box_{g_0}:=-\frac{1}{(-1+r^2)}\partial^2_t+(-1+r^2)\partial^2_r+\left(nr-\frac{(n-2)}{r}\right)\partial_r+\frac{\Delta_{\Sigma_{n-2}}}{r^2},
\end{equation}
where $\Delta_{\Sigma_{n-2}}$ is the Laplacian on the unit $(n-2)$-dimensional hyperboloid. Furthermore, $\mu^2:= m^2 + \xi R$ is an effective mass term combining the mass $m^2\geq 0$ together with the scalar curvature $R=-n(n-1)$ built out of $g_0$ and with the coupling parameter $\xi\in\mathbb{R}$.

Taking into account the symmetries of $(\mathcal{M},g_0)$, it is convenient to work with a mode expansion and consider the ansatz
	\begin{equation}
		\label{eq:ansatz solution psi of wave equation}
			\Psi_\ell(t,r,\theta,\varphi_1,...,\varphi_{n-3})=e^{-i\omega t}R_{\ell,\omega}(r)Y_\ell(\overline{\theta}),
	\end{equation}
where $\bar{\theta}:=(\theta,\varphi_1,...,\varphi_{n-3})$ and $Y_\ell(\bar{\theta})$ is an eigenfunction of $\Delta_{\Sigma_{n-2}}$,
$$\Delta_{\Sigma_{n-2}}Y_\ell(\bar{\theta})=\lambda_l Y_l(\bar{\theta}),$$
with eigenvalue $\lambda_\ell=-\left(\frac{n-3}{2}\right)^2-\ell^2$, $\ell\in\mathbb{R}$, {\it cf.} \cite{Limic}. Note that the spectrum of the Laplacian operator is continuous due to the hyperbolic nature of $\Sigma_{n-2}$.



\subsection{The radial equation as a hypergeometric equation}
	\label{sec:The radial equation as a hypergeometric equation}


The ansatz \eqref{eq:ansatz solution psi of wave equation} solves Equation \eqref{eq:dalembertian for the massless bh} if and only if the function $R_{\ell,\omega}(r)$ obeys the following ordinary differential equation, which we refer to as {\em radial equation}:
	\begin{equation}
		\label{eq:the radial equation wrt R of r}
			\Big\{(-1+r^2)\partial^2_r + \left( nr - \frac{(n-2)}{r} \right)\partial_r   + \frac{\lambda_\ell}{r^2} +\frac{\omega^2}{(-1 + r^2)}-\mu^2\Big\} R_{\ell,\omega}(r) =0.
	\end{equation}


Following \cite{Zettl:2005}, we recall that any second order differential equation can be written in Sturm-Liouville form. In the case in hand, first performing the coordinate change $r\mapsto z = \frac{-1+r^2}{r^2}\in (0,1)$, it yields
\begin{subequations}
	\label{eq:radial equation as Sturm-Liuville problem}
	\begin{equation}
		LR_{\ell,\omega}(z)=-\frac{1}{s(z)}\left(\frac{d}{dz}\left( p(z) \frac{d}{dz} \right) + q(z)  \right)R_{\ell,\omega}(z) = \omega^2 R_{\ell,\omega}(z),
	\end{equation}
	where the coefficients are
	\begin{align}
		&p(z):= z(1-z)^{\frac{3-n}{2}}, \label{eq:p(z)}\\
		&q(z):= -\frac{(1-z)^{\frac{1-n}{2}}}{4} \left(\lambda_\ell + \frac{\mu^2}{1-z}\right),\label{eq:q(z)}\\
		&s(z):=\frac{(1-z)^{\frac{1-n}{2}}}{4z}.\label{eq:s(z)}
	\end{align}
\end{subequations}

Using Frobenius method to study the asymptotic behavior of the solutions, we obtain that as $z\to 0^+$, $R_{\ell,\omega}(z)\sim z^{\alpha_\pm}$, while as $z\to 1^-$, $R_{\ell,\omega}(z)\sim (1-z)^{\beta_\pm}$, where
	\begin{subequations}
		\begin{align}
			\label{eq:auxiliary parameters alpha and beta plus and minus}
				&\alpha_\pm= \pm \frac{i\omega}{2},\\
				&\beta_\pm = \frac{(n-1)}{4}\pm \frac{1}{2}\sqrt{\frac{(n-1)^2}{4} + \mu^2}.
		\end{align}
	\end{subequations}
Choosing the ansatz
	\begin{equation}
		\label{eq:ansatz R(z)}
		R_{\ell,\omega}(z)=z^{\alpha_+} (1-z)^{\beta_+}h(z),
	\end{equation}
we find that $R_{\ell,\omega}(z)$ satisfies Equation \eqref{eq:the radial equation wrt R of r} if and only if $h(z)$ is in turn a solution of the Gauss hypergeometric equation
	\begin{align}
		\label{eq:general form of hypergeometric equation}
			z(1-z)h''(z)  + \left( c- z(a+b+1)  \right)h'(z)  - a b h(z) =0,
	\end{align}
with parameters
	\begin{subequations}
		\label{eq:parameters a b and c of hypergeometric equation}
		\begin{align}
			&a = \frac{1}{2}\left( 1 +\nu + i(\omega + \ell)\right),\label{eq:parameters a of hypergeometric equation}\\
			&b = \frac{1}{2}\left( 1 +\nu + i(\omega - \ell)\right),\label{eq:parameters b of hypergeometric equation}\\
			&c = 1 + i\omega.\label{eq:parameters c of hypergeometric equation}
		\end{align}
The auxiliary parameter introduced above, defined as
\begin{align}
	\label{eq:auxiliary parameter nu}
		&\nu :=  \sqrt{\frac{(n-1)^2}{4} + \mu^2},
\end{align}
\end{subequations}
can assume values in $(0, \infty)$ if one imposes the Brei\-ten\-lohn\-er-Freed\-man bound on the effective mass, namely $\mu^2>-\frac{(n-1)^2}{4}$, which represents a standard mode-stability requirement \cite{Breitenlohner:1982jf}.

It is also worth mentioning that the parameters of the hypergeometric equation given by Equations \eqref{eq:parameters a b and c of hypergeometric equation} assume these particular values in the case $r_h=L=1$. If we consider a massless hyperbolic black hole with a horizon at $r= r_h$, which corresponds to taking $f(r)=\frac{r^2-r_h^2}{L^2}$ in Equation \eqref{Eq: metric}, these parameters, as well $\alpha_\pm$ and $\beta_\pm$, would show a dependence on $r_h$ and in $L$. In addition, the parameters $a$ and $b$ would also depend on the spacetime dimension, namely in place of the factor $\ell$ one would have $\frac{1}{2 i r_h} \sqrt{(n-3)^2(r_h^2-L^2 ) - 4 L^2 \ell^2 }$.


\subsection{Square-integrable solutions and boundary conditions}
	\label{sec:The square-integrable solutions of the radial equation}


Following the rationale behind Sturm-Liouville theory, {\it cf.} \cite{Zettl:2005}, our next step consists of classifying the solutions of Equation \eqref{eq:the radial equation wrt R of r} and \eqref{eq:general form of hypergeometric equation} in terms of their square-integrability close to the endpoints $z\in\{0,1\}$. In practical terms, we shall look for a basis of solutions of Equation \eqref{eq:the radial equation wrt R of r} and we shall individuate the elements which lie either in $L^2((0,z_0);s(z)dz)$ or in $L^2((z_1,1);s(z)dz)$ where $z_0,z_1\in(0,1)$ can be arbitrarily chosen, while $s(z)$ is defined in Equation \eqref{eq:s(z)}.
Similarly to \cite{Bussola:2017wki,Dappiaggi:2018xvw} and using the classification of the hypergeometric equations \eqref{eq:general form of hypergeometric equation} in terms of the parameters in Equation \eqref{eq:parameters a b and c of hypergeometric equation}, see {\it e.g.} \cite{higher}, we shall highlight two regimes depending on the values assumed by the auxiliary parameter $\nu$, {\it cf.} Equation \eqref{eq:auxiliary parameter nu}. Hence, starting from the ansatz \eqref{eq:ansatz R(z)}, we distinguish between two scenarios

\begin{enumerate}
	\item[i.] \underline{Neither $c$ nor $c-a-b$ is integer valued}\\
	A convenient basis of solutions to analyze Equation \eqref{eq:the radial equation wrt R of r} close to $z=0$ is
		\begin{subequations}
		\label{eq:sol0hypergeo}
			\begin{align}
				&\mathcal{R}_{1(0)}=z^{\alpha_+}(1-z)^{\beta_+}F(a,b;c;z),\label{eq:sol10hypergeo}\\
				&\mathcal{R}_{2(0)}=z^{-\alpha_+}(1-z)^{\beta_+}F(a-c+1,b-c+1;2-c;z),\label{eq:sol20hypergeo}
			\end{align}
		\end{subequations}
where, for simplicity of notation, we have dropped the subscripts $\omega,\ell$. Taking into account the measure ruled by $s(z)$, {\it cf.} Equation \eqref{eq:s(z)}, we observe that if $\omega\in\mathbb{R}$ none of the solutions lies in $L^2((0,z_0);s(z)dz)$, while if $\mathrm{Im}(\omega)<0$ ({\em resp.} $\mathrm{Im}(\omega)>0$) then $\mathcal{R}_{1(0)}(z)$ ({\em resp.} $\mathcal{R}_{2(0)})$ lies in $L^2((0,z_0);s(z)dz)$. The behaviour of the solution for complex values of $\omega$ plays a key role in the construction of the fundamental solutions associated to the Klein-Gordon operator $P$.

Focusing instead on the endpoint $z=1$, a convenient basis of solutions is
\begin{subequations}
	\label{eq:sol1hypergeo}
	\begin{align}
		&\mathcal{R}_{1(1)}=z^{\alpha_+}(1-z)^{\beta_+}F(a,b;a+b+1-c;1-z),\label{eq:sol11hypergeo}\\
		&\mathcal{R}_{2(1)}=z^{\alpha_+}(1-z)^{\beta_-}F(c-a,c-b;c-a-b+1;1-z).\label{eq:sol21hypergeo}
	\end{align}
\end{subequations}
where again, for simplicity of notation, we have dropped the subscripts $\omega,\ell$. Analogously, considering the measure ruled by $s(z)$, we observe that $\mathcal{R}_{1(1)}\in L^2((z_1,1);s(z)dz)$ for all admissible values of $\nu$, but $\mathcal{R}_{2(1)}\in L^2((z_1,1);s(z)dz)$ if and only if $\nu<1$. In the language of Sturm-Liouville theory, {\it cf.} \cite{Zettl:2005}, $\mathcal{R}_{1(1)}$ is also referred to as principal solution, so to highlight its distinguished role.
As a matter of fact, $\mathcal{R}_{1(1)}$ is the unique solution of Equation \eqref{eq:the radial equation wrt R of r} up to scalar multiples such that $\lim\limits_{z\to 1} \mathcal{R}_{1(1)}(z)/\mathcal{R}(z)=0$ where $\mathcal{R}$ is any solution of Equation \eqref{eq:the radial equation wrt R of r} which is not a scalar multiple of $\mathcal{R}_{1(1)}$.

\item[ii.] \underline{Otherwise}\\
If $c$ is integer-valued, then the two solutions \eqref{eq:sol0hypergeo} are not linearly independent. In view of Equation \eqref{eq:parameters c of hypergeometric equation}, this occurs if and only if $\omega\in i \mathbb{Z}$. In this case, one needs to introduce an additional solution to identify a basis, {\it cf.} \cite{higher}. Yet, we omit reporting it explicitly since it is not square integrable in a neighbourhood of $z=0$ and thus we shall not use it in the rest of the paper.

If $c-a-b$ is integer-valued, this entails that $\nu\in\mathbb{N}_0$, {\it cf.} Equation \eqref{eq:auxiliary parameter nu}. In this case, one has to consider another solution since $\mathcal{R}_{1(1)}$ is proportional to $\mathcal{R}_{2(1)}$. Yet, if $\nu\geq 1$, only $\mathcal{R}_{2(1)}\in L^2((0,z_0);s(z)dz)$ is square-integrable, hence there is no need to report explicitly the new solution. An interested reader can still refer to \cite{higher}.
\end{enumerate}

Following the same strategy as in \cite{Dappiaggi:2016fwc,Bussola:2017wki,Dappiaggi:2018xvw}, when both elements of the basis of solutions are square-integrable close to an endpoint, it is necessary to impose there a boundary condition of Robin type to single out a unique representative. In the case in hand, this occurs at $z=1$ when $0<\nu<1$. Concretely, this translates to the statement that a solution $\mathcal{R}_{\gamma}$ of \eqref{eq:the radial equation wrt R of r} satisfies a {\em Robin boundary condition} at $z=1$, parameterized by $\gamma\in [0,\pi)$, if
\begin{equation} \label{eq:RBC}
	\lim_{z \to 1} \left\{ \cos(\gamma) \mathcal{W}_z[\mathcal{R}_{\gamma}, \mathcal{R}_{1(1)}] + \sin(\gamma) \mathcal{W}_z[\mathcal{R}_{\gamma}, \mathcal{R}_{2(1)}] \right\} = 0 \, ,
\end{equation}
where $\mathcal{R}_{1(1)}, \mathcal{R}_{2(1)}$ are defined in Equation \eqref{eq:sol1hypergeo}. Here, $\mathcal{W}_z[u,v] \doteq u(z)v'(z) - v(z)u'(z)$ is the Wronskian computed with respect to two differentiable functions $u$ and $v$. As a consequence, the solution $\mathcal{R}_{\gamma}$ may be written as
\begin{equation}\label{eq:Robin_solution}
	\mathcal{R}_{\gamma}(z) = \cos(\gamma)\mathcal{R}_{1(1)}(z)+\sin(\gamma)\mathcal{R}_{2(1)}(z) \, .
\end{equation}

We note that $\gamma=0$ corresponds to the standard Dirichlet boundary condition since it guarantees that $\mathcal{R}_{\gamma}$ coincides with the principal solution $\mathcal{R}_{1(1)}$. At the same time, if $\gamma=\frac{\pi}{2}$, we say that $\mathcal{R}_{\gamma}$ satisfies a Neumann boundary condition, coinciding with $\mathcal{R}_{2(1)}$. Yet, contrary to the Dirichlet boundary condition, this is not a universal assignment as it depends on the choice of the non principal solution $\mathcal{R}_{2(1)}$.

Observe that Robin boundary conditions can only be applied when $0<\nu<1$, as analyzed in the last section. If $\nu>1$, only the principal solution $\mathcal{R}_{1(1)}$ is square-integrable in a neighborhood of $z=1$ and, hence, no boundary condition is required. In practice, this is as if the Dirichlet boundary condition had been chosen.

A similar reasoning could be applied at $z=0$, but, as we have shown in the preceding subsection, if we focus only on square integrable solutions, only one exists, provided that $\Imag[\omega] \neq 0$. Therefore, at $z=0$ there is no need to impose any boundary condition.


\section{Ground and KMS states with Robin boundary conditions}
	\label{sec:Physically Sensible States}


In this section, we outline the construction of a ground state and of thermal/KMS states in the case under investigation. We start from the former and we follow the same rationale used in \cite{Dappiaggi:2016fwc,Dappiaggi:2018pju,Dappiaggi:2018xvw,Bussola:2017wki}, to which we refer for further details.

More precisely, we are interested in building a bidistribution $\omega_2\in\mathcal{D}^\prime(\mathcal{M}\times\mathcal{M})$, dubbed two-point function, such that
 \begin{align}
	 &(P\otimes\mathbb{I})\omega_2=(\mathbb{I}\otimes P)\omega_2=0,   \\
	 &\omega_2(f,f)\geq0,\, \forall f\in C_0^\infty (\mathcal{M}),
 \end{align}
where $P$ is the Klein-Gordon operator, {\it cf.} Equation \eqref{eq:KG equation}. In addition, we impose that the antisymmetric part of $\omega_2$ coincides with the so-called causal propagator $E$, which is the difference between the advanced and retarded fundamental solutions associated to $P$, {\it i.e.}, working at the level of integral kernels
\begin{equation}
	\label{eq:ccr0}
	iE(x,x')=\omega_2(x,x')-\omega_2(x',x)\text{ for }x,x'\in \mathcal{M}.
\end{equation}
The causal propagator $E$ is also the building block to implement, covariantly, the canonical commutation relations in the quantization of the underlying massive, real scalar field. In the case in hand, it can be determined as a solution of the following initial value problem, see {\it e.g.} \cite{Dappiaggi:2016fwc}
	\begin{subequations}
		\label{eq:ccrs}
		\begin{align}
			&P_xE(x,x^\prime)=P_{x^\prime}E(x,x^\prime)=0\\
			&E(x,x')|_{t'=t}  = 0 \text{,}\label{eq:ccr1}\\
			&\partial_t E(x,x')|_{t=t'}=-\partial_{t'} E(x,x')|_{t'=t} = \frac{\delta(z-z^\prime)\delta(\bar{\theta}-\bar{\theta}^\prime)}{s(z)}\label{eq:ccr2}\text{,}
		\end{align}
	\end{subequations}
where the subscripts $x$ and $x^\prime$ refer to the variable with respect to which the Klein-Gordon operator $P$ acts. In addition, $s(z)$ is the same function as in Equation \eqref{eq:s(z)}, while $\delta(\bar{\theta}-\bar{\theta}^\prime)$ denotes the product between $(n-2)$ delta distributions, one for each angular coordinate.

Our next step consists of finding an explicit expression for $E$. This is possible exploiting the isometries of the underlying metric, {\it cf.} Equation \eqref{Eq: metric}. To start with, in view of the invariance under time translation and of $\Sigma_{n-2}$ being an homogeneous manifold, we can write
	\begin{equation}
		\label{eq:ansatzTWO}
		E(x,x')=i\lim_{\varepsilon\rightarrow 0^+} \int_{\mathbb{R}} d\ell\int_{\mathbb{R}} d\omega \frac{\sin(-i\omega (t-t'-i\varepsilon))}{\omega} Y_\ell(\bar{\theta})Y_\ell(\bar{\theta}')\widetilde{E}_{2}(z,z')\text{,}
	\end{equation}
where $Y_\ell(\overline{\theta})$ are the hyperbolic harmonics, as in Equation \eqref{eq:ansatz solution psi of wave equation}, whose explicit expression, together with their completeness relation is reported in Appendix \ref{sec:The Laplace operator on the hyperbolic space}.

To impose the initial conditions we observe that, if $\widetilde{E}_2(z,z^\prime)=\widetilde{E}_2(z^\prime,z) $, then Equation (\ref{eq:ccr1}) holds true. At the same time, Equation \eqref{eq:ccr2} entails
\begin{align}
	\label{eq:adefdfiosdjjd}
		\partial_t E(x,x^\prime)|_{t=t^\prime}=\lim_{\varepsilon\rightarrow 0^+} \int_{\mathbb{R}} d\ell\int_{\mathbb{R}} d\omega Y_\ell(\bar{\theta})Y_\ell(\bar{\theta}')\widetilde{E}_{2}(z,z') \cosh(\omega\epsilon).
\end{align}
In view of the completeness relation of the hyperbolic harmonics, Equation (\ref{eq:ccr2}) amounts to imposing
	\begin{equation}
		\label{eq:ccr2 implies}
			\int_{\mathbb{R}} d\omega\widetilde{E}_2(z,z^\prime) = \frac{\delta(z-z^\prime)}{s(z)} \text{.}
	\end{equation}

In other words, the only unknown part of the causal propagator is the radial part $\widetilde{E}_2(z,z^\prime)$, and in the following we will use the same procedure employed in \cite{Dappiaggi:2016fwc,Dappiaggi:2018xvw,Bussola:2017wki} to construct it from Equation \eqref{eq:ccr2 implies} using spectral techniques. In addition, it will turn out that this analysis suffices also to write down both the ground and the KMS states for the case in hand.


\subsection{The radial Green function on the complex plane}
	\label{sec: The radial Green function on the complex plane}


The first step to write down $\widetilde{E}_2(z,z^\prime)$ consists of writing the Green function $G(z,z',\omega)$ of the radial equation, which we call {\em radial Green function}, {\it cf.} \cite[Ch. 4]{gerlach2009linear}. In view of the analysis in Section \ref{sec:The square-integrable solutions of the radial equation} of the square-integrable solutions of the radial equation \eqref{eq:radial equation as Sturm-Liuville problem}, we split $G$ in two parts,
	\begin{equation*}
		G(z,z',\omega)=\begin{cases}G_{\mini{<}}(z,z',\omega)\text{, for Im}(\omega)<0,\\ G_{\mini{>}}(z,z',\omega)\text{, for Im}(\omega)>0, \end{cases}
	\end{equation*}
which are connected to each other due to the symmetries of the radial solutions as follows $$G_{\mini{>}}(z,z',\omega)=G_{\mini{<}}(z,z',\overline{\omega})=\overline{G_{\mini{<}}(z,z',\omega)} .$$ For this reason, we focus only on $\Imag(\omega)<0$. In addition notice that the case $\Imag(\omega)=0$ is left out since no square-integrable solution exists in this scenario close to the endpoint $z=0$. Explicitly, we have
	\begin{align*}
		&G_{\mini{<}}(z,z';\omega)=\frac{1}{\mathcal{N}_{\mini{<}}}\left\{\Theta(z-z')R_{1(0)}(z)R_\gamma(z')+\Theta(z'-z)R_{1(0)}(z')R_\gamma(z)\right\}.
	\end{align*}
The normalization constant $\mathcal{N}_{\mini{<}}$, determined by the identity $LG=\delta$, is
\begin{equation}
	\mathcal{N}_{\mini{<}} :=  p(z)	\mathcal{W}\{R_{1(0)},R_{\gamma}\},
\end{equation}
where $p(z)$ is given by Equation \eqref{eq:p(z)} while $\mathcal{W}$ is the Wronskian between the solutions $R_{1(0)}$ and $R_{\gamma}$. Using the connection formulae for the hypergeometric functions, {\it cf.} Equations (15.10.5) and (15.10.21) from \cite[Ch. 15]{NIST}, we find that the normalization constant reads
	\begin{align}
		\mathcal{N}_{\mini{<}}=-\nu(A\sin(\gamma)-B\cos(\gamma)),
	\end{align}
where, recalling the coefficients in Equation \eqref{eq:parameters a b and c of hypergeometric equation},
  \begin{subequations}
			\label{eq:CoefAeB}
		\begin{align}
			A&=\frac{\Gamma(c)\Gamma(c-a-b)}{\Gamma(c-a)\Gamma(c-b)}, \label{eq:CoefA}\\
			B&=\frac{\Gamma(c)\Gamma(a+b-c)}{\Gamma(a)\Gamma(b)}. \label{eq:CoefB}
		\end{align}
	\end{subequations}
Observe that $\overline{A(\omega)}=A(\overline{\omega})$ and $\overline{B(\omega)}=B(\overline{\omega})$ if  $\mathrm{Re}(\omega)=0$, while  $\overline{A(\omega)}=A(-\omega)$ and $\overline{B(\omega)}=B(-\omega)$ if $\Imag(\omega)=0$.


\subsection{The poles of the radial Green function}
	\label{sec: The poles of the radial Green function}


Recalling that our goal is the construction of $\widetilde{E}_2(z,z^\prime)$, and hence of the causal propagator as in Equation \eqref{eq:ansatzTWO}, the next step consists of finding out whether the radial Green function possesses poles or not. As a matter of fact, the existence of poles depends on the boundary condition chosen through the parameter $\gamma\in[0,\pi)$, {\it cf.} Equation \eqref{eq:RBC}. Moreover, these poles are nothing but the zeros of the normalization constant seen as a function of the frequency, which in turn correspond to the degenerate cases of the hypergeometric solutions. 

To start with, we focus on Dirichlet and Neumann boundary conditions. In these cases the zeros of $\mathcal{N}_{\mini{<}}$ are respectively
\begin{subequations}
\begin{align}
	\label{eq:psodklpsodlp}
		&B=0\iff\omega_{\mini{D}} = \pm \ell + i (2p + 1 + \nu),\\
		&A=0\iff\omega_{\mini{N}} = \pm \ell + i (2p + 1 - \nu), \text{ for }p\in\mathbb{N}\cup\{0\}.
\end{align}
\end{subequations}

The above set of poles coincides with the quasinormal modes. For the case of Dirichlet boundary condition, it is consistent with the quasinormal modes obtained in \cite{Aros:2002te}. Note that the poles lying on the upper or lower part of the complex-plane is a matter of convention, consequence of choosing $\alpha_+$, instead of $\alpha_-$ as in \cite{Aros:2002te}. The smallest values for the imaginary part of the poles above are
\begin{subequations}
\begin{align}
	\label{eq:psodklpsodlp2}
	&\Imag(\omega_{\mini{D}}^{\mini{S}}) = + (1 + \nu),\\
	&\Imag(\omega_{\mini{N}}^{\mini{S}}) = + (1 - \nu).
\end{align}
\end{subequations}
For $\nu\in(0,1)$, all poles lie on the upper part of the complex plane, hence $\mathcal{N}_{\mini{<}}$ has no zeros and the Green function has no poles for Dirichlet or Neumann boundary condition. Recall that for $\nu>1$ only Dirichlet boundary condition is applicable.

Secondly, let us consider Robin boundary conditions. In this case, we need to invoke Cauchy's argument principle. Here goes a qualitative description for the cases considered here, when there are no poles on the lower part of the complex-plane.
Define $\Xi(\omega) := \frac{B}{A}$. The zeros of the function
\begin{equation}
	H(\omega):= \tan(\gamma) - \Xi(\omega)
\end{equation}
correspond to the poles of the radial Green function. The poles of $H$ are just those of $\Xi$, which coincide with $\omega_{\mini{N}}$. Note that $\gamma =\arctan(\Xi(\omega))$ exists whenever $\Xi(\omega)$ is real and this only happens for purely imaginary frequencies. Therefore, the region where $H(\omega)$ always has a zero is the half-line on the negative $\Imag(\omega)$-axis stretching, to negative infinity, from the pole with smallest imaginary part, say $\omega_{0}^{\mini{S}}$. This region corresponds to $\gamma\in[\gamma_0,\gamma_\infty]$ for $\gamma_0:=\arctan(\Xi(\omega_0^{\mini{S}}))$ and  $\gamma_\infty:=\lim\limits_{\omega\rightarrow\infty}\arctan(\Xi(-i\omega))$. We choose positive values for $\gamma_0$ and $\gamma_\infty$, and we note that $\gamma_0\in\left(\frac{\pi}{2},\pi\right)$ for $\nu\in(0,1)$.

Let $\mathcal{Z}$ and $\mathcal{P}$ be the number of zeros and poles of $H$. Invoking the argument principle, and knowing the poles of $H$, we just have to check the stability of a sequence of contour integrals on $C_n$ exhausting the lower half of the $\omega$-complex plane
\begin{equation}
	\oint_C \frac{H'(z)}{H(z)}dz=2\pi i (\mathcal{Z}-\mathcal{P}).
\end{equation}
$\mathcal{P}$ equals the number of poles of $\Xi$, so, for each case, we can compute the contour integrals and check that $\mathcal{Z}$ stabilizes to $1$. Therefore, for $\omega$ along the semi-axis Im$(\omega)<0$, there exists one and only one pole. By symmetry, we know that its complex conjugate is also a pole for $G(z,z';\omega)$.

In short, in the $\omega^2$-complex plane, the radial Green function $G(z,z';\omega)$ has a branch cut on the positive real line, but for $\gamma\in [0,\gamma_0)$ it has no poles.


\subsection{The resolution of the identity}
	\label{sec: The spectral resolution}

\begin{figure}[H]
\centering
	\includegraphics[width=.8\textwidth]{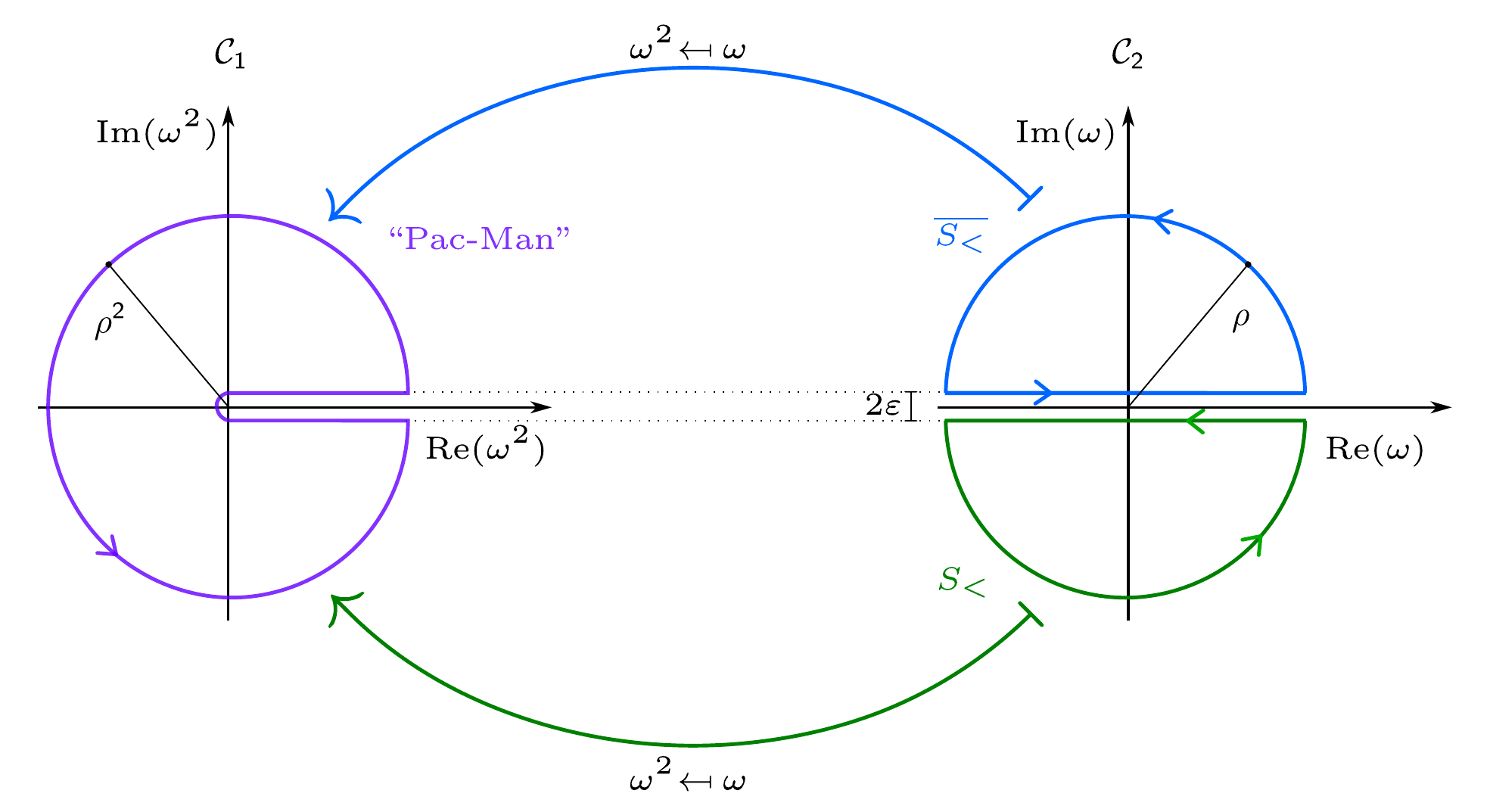}
\caption{Contours $\mathcal{C}_1$ and $\mathcal{C}_2$ of Equation \eqref{eq:contourIntegrals}. Each semi-circle on the right is mapped to a ``Pac-Man'' contour. Since we take $\mathcal{C}_2$ to be both semi-circles, the factor $2$ coming from $d\omega^2=2\omega d\omega$ gets canceled}
\label{fig:contour}
\end{figure}

Having constructed the radial Green function in Section \ref{sec: The radial Green function on the complex plane} and having analyzed its simple poles in Section \ref{sec: The poles of the radial Green function}, we have all ingredients to derive an explicit form for $\widetilde{E}_2(z,z^\prime)$ via Equation \eqref{eq:ccr2 implies}. More precisely, it holds, {\it cf.} \cite{greenBook}

	\begin{equation}
		\label{eq:identity radial green function}
			\frac{1}{2\pi i}\oint_{\mathcal{C}^\infty} d(\omega^2) G(z,z';\omega) = -\frac{\delta(z-z')}{s(z)} \text{,}
	\end{equation}
where we have promoted the spectral parameter $\omega^2$ to a complex coordinate and where $\mathcal{C}^\infty$ denotes a circle of infinite radius in the $\omega^2$-plane with a counterclockwise orientation. Focusing on the class of Robin boundary conditions,  $\gamma\in(0,\gamma_0]$, for which the radial Green function has no poles,
it holds
	\begin{equation}
		\label{eq:contourIntegrals}
		\oint_{\mathcal{C}_1} d(\omega^2) G(z,z';\omega) = \oint_{\mathcal{C}_2} d\omega\omega G(z,z';\omega),
	\end{equation}
where $\mathcal{C}_1$ and $\mathcal{C}_2$ are like in Figure \ref{fig:contour}.

Let $G^{\mini{R/L}}$ denote the right and left propagating components of the radial Green function $G$, i.e.
	\begin{align}
		&G^{\mini{R}}_{\mini{<}}(z,z'):=\mathcal{N}_{\mini{<}}^{\mini{-1}}\Theta(z-z')R_{1(0)}(z)R_\gamma(z'),
	\end{align}
while $G^{\mini{L}}_{\mini{<}}(z,z'):=G(z,z^\prime)- G^{\mini{R}}_{\mini{<}}(z,z') $. Then the  contour integral in the $\omega$-plane can be written as
	\begin{align}
		\oint_{C_2^\infty} d\omega\omega G(z,z';\omega)=\lim\limits_{\varepsilon\rightarrow0}\lim\limits_{\rho\rightarrow\infty} \left\{\oint_{\overline{S_{\mini{<}}}} d\omega\omega \overline{G_{\mini{<}}(z,z';\omega)} +\oint_{S_{\mini{<}}} d\omega\omega G_{\mini{<}}(z,z';\omega)\right\}.
	\end{align}
Focusing separately on the left and right propagating components, say on $G^{\mini{R}}$, we can first deal with the limit as $\rho$ diverges. Since only a finite number of simple poles occurs, one can use the same argument as in \cite[Appendix A]{Bussola:2017wki} to infer that the limit for large values of $\rho$ reduces to
	\begin{align}
		\oint_{C_2^\infty} d\omega&\omega G^{\mini{R}}(z,z';\omega)=  \Theta(z-z')\lim\limits_{\varepsilon\rightarrow0}   \int_{\mathbb{R}} dr \Delta G^{\mini{R}};
	\end{align}
where
	\begin{align}
		\Delta G^{\mini{R}}=(r+i\varepsilon)\overline{\left( \frac{ R_{1(0)}(z)R_\gamma(z')  }{-\nu(A\sin(\gamma)-B\cos(\gamma))} \right)}\Bigg|_{\omega=r+i\varepsilon}-\textrm{c.c.},
	\end{align}
where $\textrm{c.c.}$ stands for the complex conjugate. At last taking the limit as $\epsilon\to 0$, we find
	\begin{align}
		\label{eq:introducetheta}
		 \oint_{C_2^\infty} d\omega&\omega G^{\mini{R}}(z,z';\omega)=  \Theta(z-z')\int_{\mathbb{R}} d\omega \frac{\omega}{\nu}\frac{(\overline{A}B-A\overline{B})}{ |A\sin(\gamma) -B\cos(\gamma)|^2}   R_\gamma(z)R_\gamma(z') .
	\end{align}
The integrand on the right hand side of Equation (\ref{eq:introducetheta}) is symmetric under the exchange $z\leftrightarrow z'$ and it is even in $\omega$. Hence we can combine the contribution from the right and from the left propagating Green function obtaining a so-called resolution of the identity in terms of eigenfunctions of the operator $L$, {\it cf.} Equation \eqref{eq:radial equation as Sturm-Liuville problem} with Robin boundary conditions parameterized by $\gamma$
\begin{align}
	\label{eq:leftandrighttogether}
	 \frac{\delta(z-z')}{s(z)} = -\frac{1}{2\pi i} 
	 \int_{\mathbb{R}} d\omega \frac{\omega}{\nu}\frac{(\overline{A}B-A\overline{B})}{ |A\sin(\gamma) -B\cos(\gamma)|^2} R_\gamma(z)R_\gamma(z').
\end{align}
To conclude, combining Equation (\ref{eq:ccr2 implies}), (\ref{eq:identity radial green function}) with Equation (\ref{eq:introducetheta}), we can infer
	\begin{align}
		\label{eq:radial green on positive frequencies}
		\boxed{\widetilde{E}_2(z,z')  =- \omega \mathcal{N}^{\mini{-1}} R_\gamma(z)R_\gamma(z')},
	\end{align}
	where
	\begin{align}\label{eq:normwithXi}
		& \mathcal{N}^{\mini{-1}}:=\frac{1}{2 \pi i \nu}\frac{  (\overline{A}B-A\overline{B})}{ |A\sin(\gamma) -B\cos(\gamma)|^2} = \frac{1}{ \pi \nu}\frac{\Imag(\Xi) }{\cos(\gamma)^2 |\tan(\gamma) - \Xi|^2 }\in \mathbb{R}.
	\end{align}

\subsection{Construction of the ground and of the KMS states}
	\label{sec: A ground state and a KMS state}


In this section, we shall start from the explicit expression of the causal propagator $E$, {\it cf.} Equation \eqref{eq:ansatzTWO} and \eqref{eq:radial green on positive frequencies} to construct a class of physically sensible two-point functions. As discussed at the beginning of Section \ref{sec:Physically Sensible States} these are positive distributions on $\mathcal{M}\times\mathcal{M}$ which fulfill the equation of motion and whose antisymmetric part coincides with the causal propagator. In between the plethora of admissible two-point functions, it is universally accepted that the so-called Hadamard condition selects those which are physically sensible. Here we shall not enter into a detailed analysis of its definition, referring an interested reader to \cite{Khavkine:2014mta} for an extensive review and to \cite{Dappiaggi:2017wvj,Wrochna:2016ruq} for an analysis of the case of asymptotically AdS spacetimes.

As proven in \cite{sahlmann2000passivity}, one notable class of two-point functions obeying the Hadamard condition is the one associated to a ground state. This state can only exist if the underlying spacetime is stationary, as in the case of our interest and it is completely specified by the requirement that the two-point function is built only out of positive frequencies. In other words, starting from Equation \eqref{eq:ansatzTWO} and \eqref{eq:radial green on positive frequencies}, it follows that the two-point of the ground state for a massive, real scalar field, obeying the Klein-Gordon equation with Robin boundary conditions with parameter $\gamma\in[0,\gamma_0)$ is
\begin{equation}
	\omega_2(x,x^\prime)=\lim_{\varepsilon\rightarrow 0^+} \int_{\mathbb{R}} d\ell\int_{0}^\infty d\omega e^{-i\omega(t-t^\prime-i\epsilon)}\mathcal{N}^{\mini{-1}} Y_\ell(\bar{\theta})Y_\ell(\bar{\theta}') R_\gamma(z)R_\gamma(z')\text{.}
\end{equation}
Given a ground state such as above, it is straightforward to construct a thermal/KMS state, still exploiting the translation invariance of the underlying metric along the time coordinate $t$, {\it cf.} Equation \eqref{Eq: metric}. More precisely, the KMS condition at inverse temperature $\beta=T^{-1}>0$, such as in \cite[Eq. (3.11)]{haag1984quantum}, is guaranteed to hold for a two-point function $\omega_{2,\beta}\in\mathcal{D}^\prime(\mathcal{M}\times\mathcal{M})$ whose integral kernel reads
	\begin{align}
		\label{eq:KMSstate}
			\omega_{2,\beta}(x,x^\prime)=\lim_{\varepsilon\rightarrow 0^+} \int_{\mathbb{R}} d\ell\int_{\mathbb{R}} d\omega\frac{\Theta(\omega)}{e^{\beta\omega}-1} Y_\ell(\bar{\theta})Y_\ell(\bar{\theta}')\widetilde{\omega}_{2}(z,z') \left[ e^{\beta\omega}e^{-i \omega (t-t'-i\varepsilon)}+e^{+i \omega (t-t'+i\varepsilon)} \right],
	\end{align}
where $\Theta$ is the Heaviside distribution and $\widetilde{\omega}_{2}(z,z')$ is related with $\widetilde{E}_2(z,z')$, given by Equation \eqref{eq:radial green on positive frequencies}, by the expression $\widetilde{E}_2=-\omega \,\widetilde{\omega}_2$, consequence of Equation \eqref{eq:ccr0}. One can infer that, for each value of $\beta$, the difference between $\omega_{2,\beta}$ and $\omega_2$ is smooth. This entails that each KMS state is of Hadamard form and thus physically acceptable.

To conclude the section, we highlight one notable property of the KMS states. More precisely, taking the Fourier transform of the two-point function $\omega_2$, with respect to $s=t-t'$ and calling $\Omega$ the associated momentum, yields
\begin{align}
	\label{eq:Fourier transform ground state}
		\mathcal{F}[\omega_2](\Omega)&=\lim_{\varepsilon\rightarrow 0^+} \int_{\mathbb{R}} d\ell\int_{\mathbb{R}} d\omega\Theta(\omega) Y_\ell(\bar{\theta})Y_\ell(\bar{\theta}')\widetilde{\omega}_{2}(z,z') e^{-\omega\varepsilon}\delta(\omega +\Omega).
\end{align}
Note that $\mathcal{F}[\omega_2](\Omega)$ is non-vanishing only for negative energy gap $\Omega$. On the other hand, for any KMS state one obtains
	\begin{align}
		\label{eq:Fourier transform KMS state}
			\mathcal{F}[\omega_{2,\beta}](\Omega)&=\lim_{\varepsilon\rightarrow 0^+} \int_{\mathbb{R}} d\ell\int_{\mathbb{R}} d\omega\Theta(\omega) Y_\ell(\bar{\theta})Y_\ell(\bar{\theta}')\widetilde{\omega}_{2}(z,z') \frac{e^{-\omega\varepsilon}}{e^{\beta\omega}-1}\left[ e^{\beta\omega} \delta(\omega+\Omega)+ \delta(\omega-\Omega) \right].
	\end{align}
Therefore, we have that $\mathcal{F}[\omega_{2,\beta}](\Omega)$ is defined, an not generally vanishing, for both positive and negative energy gaps. In fact, it satisfies the detailed balance condition
	\begin{equation}
		\label{eq:detailedB}
			\mathcal{F}[\omega_2](-\Omega)=e^{\beta\Omega}\mathcal{F}[\omega_2](\Omega).
	\end{equation}


\section{An Unruh-DeWitt detector}
	\label{sec:An Unruh-DeWitt detector}


In this section, consider an Unruh-DeWitt detector traveling in spacetime along a smooth timelike curve $x(\tau)$ parameterized by its proper time $\tau$. Along the lines of  \cite{DeWitt,Unruh}, we assume the detector to be a spatially localized two-level system. Hence the detector can be either in the ground state $\ket{0}$ or in an excited state $\ket{\Omega}$. The pair $\{\ket{0},\ket{\Omega}\}$ forms an orthonormal basis for a Hilbert space $\mathcal{H}_{\mini{D}}\simeq\mathbb{C}^2$ and they are eigenstates of the detector's free Hamiltonian. In other words, it holds $H_{\mini{D}}\ket{0}=0$ and $H_{\mini{D}}\ket{\Omega}=\Omega\ket{\Omega}$. Following the standard approach, see {\it e.g.} \cite{Hodgkinson:2011pc,Hodgkinson:2012mr,Hodgkinson:2014iua}, the detector can be coupled to a scalar field $\Psi$ through the interaction Hamiltonian
	\begin{equation}
		\label{eq:Hint}
			H_{int}(\tau)=c\,\Psi(x(\tau))\otimes\mu(\tau)\text{,}
	\end{equation}
where $c\in\mathbb{R}\setminus\{0\}$ is a small coupling constant while $\mu$ is the detector's monopole moment operator. The total Hilbert space of the coupled system is $\mathcal{H}_{\Psi}\otimes\mathcal{H}_{\mini{D}}$ and the total Hamiltonian is  $$H=H_\Psi\otimes \mathbb{I}_{\mini{D}}+\mathbb{I}_\Psi\otimes H_{\mini{D}}+H_{int}.$$
Within first-order perturbation theory, the transition probability of the total system from an initial state $\ket{i}=\ket{\psi,0}$ at time $\tau_i$, to a final state $\ket{f}=\ket{\phi,\Omega} $ at time $\tau_f$ is
	\begin{equation}
		|\mathcal{M}|^2=c^2|\bra{\Omega}\mu(0)\ket{0}|^2\left|\int_{\tau_i}^{\tau_f}d\tau e^{i \Omega\tau}\chi(\tau)\bra{\phi}\Psi(x(\tau))\ket{\psi}\right|^2.
	\end{equation}
The first term $c^2|\bra{\Omega}\mu(0)\ket{0}|^2$ depends on the internal details of the detector. The second one depends only on the state of the field and on the trajectory of the detector. We shall call it the detector's response function, or the transition probability, and we shall denote it with $\mathcal{F}$. Since we are interested in counting the detector's clicks, and not in how the interaction affects the field, we can sum over all final states of the field $\ket{\phi}$ and by completeness, it holds
	\begin{align}
		\label{eq:justlike33}
			\mathcal{F} &= \int_{\tau_i}^{\tau_f} d\tau \int_{\tau_i}^{\tau_f}  d\tau' e^{-i\Omega(\tau-\tau')}\bra{\psi}\Psi(x(\tau))\Psi(x(\tau'))\ket{\psi},
	\end{align}
where $\bra{\psi}\Phi(x(\tau))\Phi(x(\tau'))\ket{\psi}$ is the pullback along the detector's trajectory of the two-point function $\omega_2(x,x')$ associated to the state $\ket{\psi}$.

In the scenario where the detector is following a static trajectory, the field is invariant under time-translations and we turn on the interaction at $\tau_i=-\infty$, we can rewrite Equation \eqref{eq:justlike33} to obtain the instantaneous \textit{transition rate}
	\begin{align}
		\label{eq:transitionjustlike33}
			\dot{\mathcal{F}} &= \int_\mathbb{R}ds e^{-i\Omega s} \omega_2(s;\underline{x}),
	\end{align}
where $s$ is the proper time difference $\tau - \tau'$ while $\underline{x}$ corresponds to fixed spatial coordinates of the detector's worldline. The dot denotes the derivative along the coordinate $\tau$. By direct inspection, one can realize that $\dot{\mathcal{F}}$ is the Fourier transform of the pullback of two-point function along the detector's trajectory with the Fourier parameter evaluated at $\Omega$. We shall also call it the response of the detector. Note that, for $\Omega>0$, Equation \ref{eq:transitionjustlike33} characterizes the amplitude of the excitations of the detector, and for $\Omega<0$, of the de-excitations.

Equation (\ref{eq:transitionjustlike33}) might not hold true when the $\varepsilon$-regularization limit in the two-point function does not commute with the proper time derivative of the transition probability. This problem has been carefully studied for Hadamard states on curved spacetimes in \cite{Louko:2007mu} and, since we are considering a detector following a static trajectory on a static spacetime, this potential problem does not occur.


For the ground state and for a KMS state, the transition rate (\ref{eq:transitionjustlike33}) is given respectively by Equation \eqref{eq:Fourier transform ground state} and \eqref{eq:Fourier transform KMS state} with $\Omega\mapsto\sqrt{|g_{00}|}\Omega$.
This rescaling is a correction that comes from switching the time interval to a proper time interval $t-t'\mapsto s/\sqrt{|g_{00}|}$, where $g_{00}=-f(r)$ as in Equation \eqref{Eq: metric}. For a detector with energy gap $\Omega$, following a static trajectory $x(\tau)=(\tau,z_{\mini{D}},\bar{\theta}_{\mini{D}})$ at fixed spatial positions $z=z_{\mini{D}}$ and $\bar{\theta}=\bar{\theta}_{\mini{D}}$, the transition rate for the ground state reads
\begin{equation}
	\label{eq:transition rate for the ground state_original}
		\dot{\mathcal{F}}_{0} =  \int_{\mathbb{R}}  d\ell Y_\ell(\bar{\theta}_{\mini{D}})^2\widetilde{\omega}_{2}(z_{\mini{D}},z_{\mini{D}})\Big|_{\omega=-\sqrt{|g_{00}|}\Omega}.
\end{equation}
For the KMS state with respect to $\partial_t$ at inverse-temperature $\beta$
	\begin{equation}
		\label{eq:transition rate for the KMS state}
			\dot{\mathcal{F}}_{\beta}= \frac{\text{sign}(\Omega)}{e^{\text{sign}(\Omega)\beta \omega }-1}\int_{\mathbb{R}}  d\ell Y_\ell(\bar{\theta}_{\mini{D}})^2\widetilde{\omega}_{2}(z_{\mini{D}},z_{\mini{D}})\Big|_{\omega=+\sqrt{|g_{00}|}\left|\Omega\right|}  ,
	\end{equation}
where $\text{sign}(\Omega)$ stands for the sign function with variable $\Omega$. Note that for a KMS state at inverse-temperature $\beta=2\pi$ the detector's response satisfies the detailed balance condition \eqref{eq:detailedB} at the local Hawking temperature $T_H$, defined by the geometrical constant Hawking temperature corrected by the time-dilation effect experienced by the detector
\begin{equation}
	\label{eq:local Hawking temperature}
	T_H := \frac{1}{2\pi \sqrt{|g_{00}|}}.
\end{equation}


\section{Numerical analyses of the transition rate}
	\label{sec:Numerical analysis}

In this section, we study the transition rate of an Unruh-DeWitt detector coupled to a ground state and to a KMS state, respectively given by Equation \eqref{eq:transition rate for the ground state} and \eqref{eq:transition rate for the KMS state}, on massless hyperbolic black holes of three and four dimensions. The analyses summarized here are mainly numerical and they can be found in a Mathematica notebook available in GitHub \cite{git_unruh_deWitt_hyperbolic_bh}. Most notably, we observe that the anti-Hawking effect does not occur on the four-dimensional case, as shown in Section \ref{sec:Anti-Hawking effect}.

We start with a few analytical considerations. 
 For $\nu\in(0,1)$, one needs to impose a boundary condition at $z=1$ and by Cauchy's argument principle, we have shown in Section \ref{sec: The poles of the radial Green function} that if $\gamma\in[0,\gamma_0)$ no poles occur, where
	\begin{subequations}
		\begin{align}
			&\tan(\gamma_0)=\Xi(0) = \frac{\Gamma(\nu)}{\Gamma(-\nu)}\frac{\left|\Gamma\left(\frac{1-\nu}{2} +\frac{i\ell}{2}\right)\right|^2}{\left|\Gamma\left(\frac{1+\nu}{2} +\frac{i\ell}{2}\right)\right|^2}.\label{eq:tan gamma 0 all nu}
		\end{align}
	\end{subequations}
In particular, for the massless conformally coupled case, $\nu=1/2$, \eqref{eq:tan gamma 0 all nu} reduces to
	\begin{equation}
		\tan(\gamma_0)= -\frac{1}{2}\frac{\left|\Gamma\left(\frac{1}{4} +\frac{i\ell}{2}\right)\right|^2}{\left|\Gamma\left(\frac{3}{4} +\frac{i\ell}{2}\right)\right|^2}\in\left(\frac{\pi}{2}, \arctan\left( -\frac{1}{2}\frac{\Gamma(1/4)^2}{\Gamma(3/4)^2}\right)+\pi\right).
	\end{equation}
Let us consider $\gamma\in[0,\pi/2]$, so that no pole occurs for any $\ell$-mode and take a detector with energy gap $\Omega$ following a static trajectory with $z=z_{\mini{D}}$ and $\bar{\theta}=\bar{\theta}_{\mini{D}}$. Putting together Equation \eqref{eq:transition rate for the KMS state} and (\ref{eq:ansatzTWO}), the transition rate for the ground state reads
	\begin{align}
		\label{eq:transition rate for the ground state}
			\dot{\mathcal{F}}_0 = \int_{\mathbb{R}}  d\ell Y_\ell(\bar{\theta}_{\mini{D}})^2 \frac{1}{2 \pi i \nu}\frac{(\overline{A}B-A\overline{B})}{ |A\sin(\gamma) -B\cos(\gamma)|^2} R_\gamma(z_{\mini{D}})^2\Big|_{\omega=-\sqrt{|g_{00}|}\Omega}.
	\end{align}
Considering Equation \eqref{eq:KMSstate} instead of (\ref{eq:ansatzTWO}), we obtain the transition rate for the KMS state, here at inverse-temperature $\beta=2\pi$

		\begin{align}
			\label{eq:transitionRindlerAdS}
				\dot{\mathcal{F}}_{T_H} = \frac{\text{sign}(\Omega)}{e^{\text{sign}(\Omega)\beta \omega }-1} \int_{\mathbb{R}}  d\ell Y_l(\bar{\theta}_{\mini{D}})^2 \frac{1}{2 \pi i \nu}\frac{(\overline{A}B-A\overline{B})}{ |A\sin(\gamma) -B\cos(\gamma)|^2} R_\gamma(z_{\mini{D}})^2\Big|_{\omega=+\sqrt{|g_{00}|}|\Omega|}.
		\end{align}

A close scrutiny of Equation \eqref{eq:transition rate for the ground state} and \eqref{eq:transitionRindlerAdS} leads to the natural question of whether or not the transition rates obtained still pertain a singular, distributional nature inherited from the two-point functions of the ground and the KMS states. We do expect them to yield a finite quantity for all admissible values of the underlying parameters. However, while the integrand is in both cases finite due to the properties of the hypergeometric functions and to the lack of poles for the admitted range of $\gamma$, the integral along the spectral parameter $\ell$, might lead to divergences. Although we do not have an analytic proof of this statement for arbitrary parameters, in Appendix \ref{sec:The transition rate is a well-defined positive quantity}, we prove that the integral over $\ell$ is indeed finite for the massless conformally coupled case. It is also worth mentioning that the integrands of the transition rates are invariant under the mapping $\ell\mapsto -\ell$.
Moreover, the transition rate is a real-valued quantity, since the solutions $R_{\gamma}$ have vanishing imaginary part, as shown in Section \ref{sec:Vanishing imaginart part for Rgamma} of Appendix \ref{sec:The transition rate is a well-defined positive quantity}.

In the following sections, we check how the response function behaves with respect to the underlying parameters for the states that we have constructed for the massless conformally coupled case. Note that the response depends on the energy gap, on the boundary condition and on the position of the detector or, equivalently, on the local Hawking temperature (\ref{eq:local Hawking temperature}) by setting 	$$z_{\mini{D}}(T_{\mini{H}})=\frac{1}{1+4\pi^2T_{\mini{H}}^2}.$$
Moreover, since the radial solutions do depend on the dimension $n$ through $\beta_+=n/4$, so does the transition rate. However, before looking at what happens on massless hyperbolic black holes, it is worth recalling what occurs in Minkowski spacetime. Therefore, in Section \ref{sec:The transition rate on Minkowski spacetime}, we use the expressions for the transition rate obtained in \cite{Hodgkinson:2011pc} to review the response of a detector measuring Unruh radiation on Minkowski spacetime. Then, with our intuition refreshed, we analyze the response of the detector measuring Hawking radiation on massless hyperbolic black holes in the remaining sections. For clarity purposes, most plots have been normalized with respect to their own maximum value, since here we are only concerned with qualitative behaviors. In any case, all plots and their respective normalization factor can be found at \cite{git_unruh_deWitt_hyperbolic_bh}.

The framework established does apply to scalar fields with arbitrary mass and coupling, and to perform a numerical analysis in these cases is straightforward. We have highlighted above the details that would need attention, namely the proof that the integral in $\ell$ converges, but we expect that to be, at least numerically, direct. One should keep in mind that not all possible values of mass and coupling allow for Robin boundary conditions to be chosen, but only those for which $\nu\in(0,1)$. In \cite{git_unruh_deWitt_hyperbolic_bh}, we have implemented the expressions of the transition rates for arbitrary parameters. However, for convenience, in the numerical analyses we consider in this paper we focus on massless conformally coupled scalar field, which corresponds to $\nu=1/2$. The goal of the numerical analyses of applying the general framework established by studying and generalizing previous results concerning the manifestation of the anti-Hawking phenomenon is reached with analyzing only this case.

\subsection{The transition rate on Minkowski spacetime}
	\label{sec:The transition rate on Minkowski spacetime}


On three- and four-dimensional Minkowski spacetime, an Unruh-DeWitt detector with energy gap $\Omega$ following a Rindler trajectory with proper acceleration $a$ and interacting with the vacuum state of a free, massless, real scalar quantum field theory has the following response function \cite{Hodgkinson:2011pc}

\begin{subequations}
	\label{eq:transition rate Minkowski 3 to 6}
  \begin{tabularx}{.9\textwidth}{Xp{2cm}X}
  \begin{equation}
  		\label{eq:transition rate Minkowski 3}
      	\dot{\mathcal{F}}_{\text{Mink}_3}= \frac{1}{2} \frac{1}{e^{2\pi \Omega/a}+1},
  \end{equation}& &
  \begin{equation}
  	\label{eq:transition rate Minkowski 4}
  	\dot{\mathcal{F}}_{\text{Mink}_4}= \frac{1}{2\pi} \frac{\Omega}{e^{2\pi \Omega/a}-1},
  \end{equation}
	\end{tabularx}
\end{subequations}

\begin{figure}[H]
\centering
	\includegraphics[width=.45\textwidth]{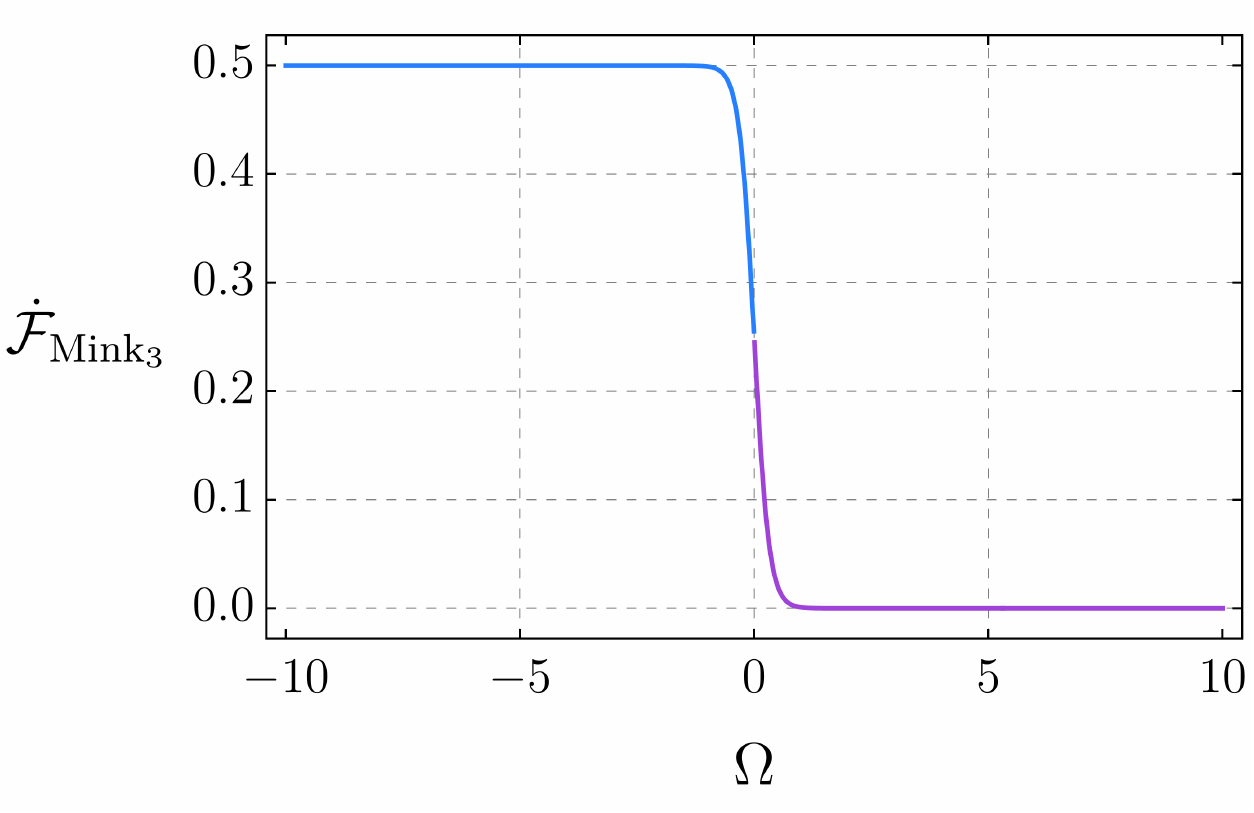}\hspace{.5cm}
		\includegraphics[width=.45\textwidth]{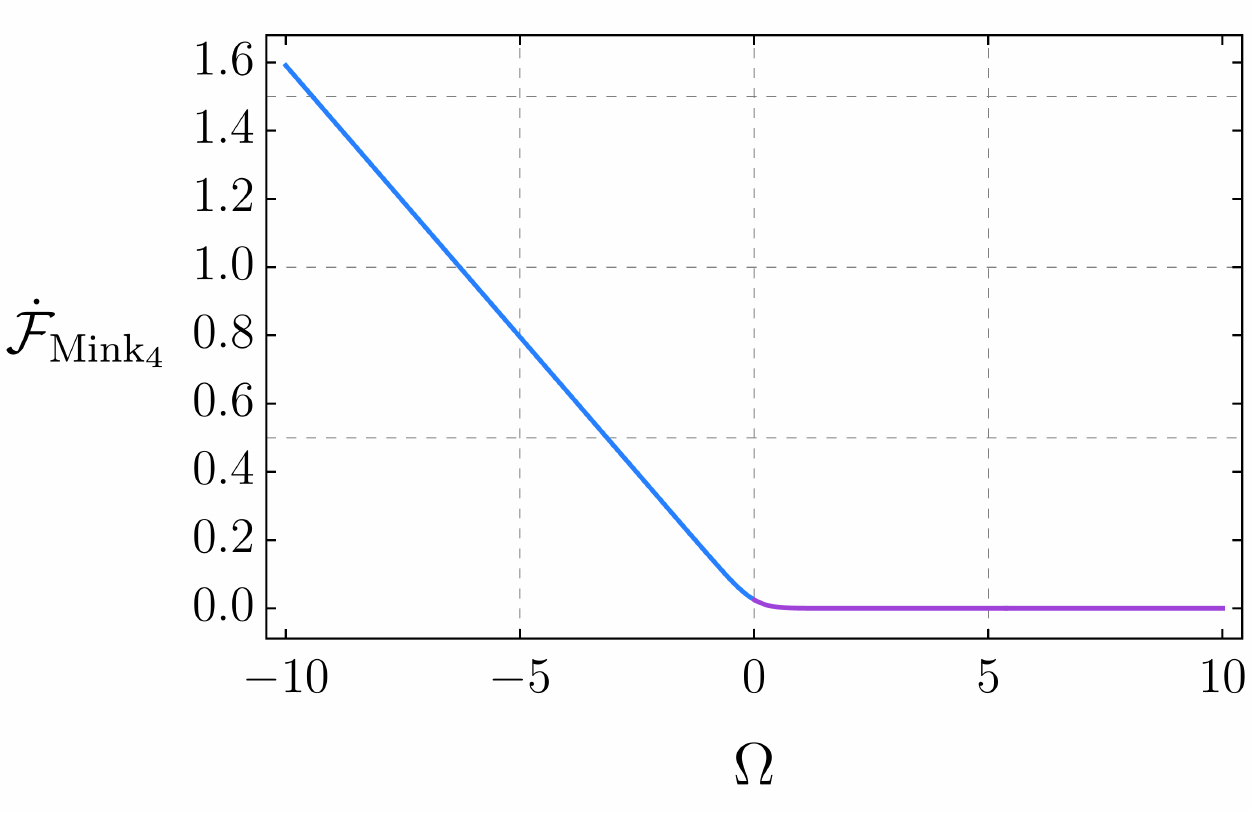}
\caption{The transition rate on Minkowski spacetime as a function of the energy gap $\Omega$ for a fixed proper acceleration $a=1$ and for spacetime dimensions three and four, from left to right.}
\label{fig:transition rate Minkowski 3 4 5 6 as a function of the energy gap}
\end{figure}
As a function of the energy gap, $\dot{\mathcal{F}}_{\text{Mink}_{3}}$ and $\dot{\mathcal{F}}_{\text{Mink}_{4}}$ behaves as expected, in the sense that, heuristically speaking it is always easier to de-excite than to excite: the lower the energy gap, the higher the transition rate, as shown in the following plot, {\it cf.} Figure \ref{fig:transition rate Minkowski 3 4 5 6 as a function of the energy gap}. On Minkowski spacetime of five and six dimensions, the qualitative behavior is the same as the one for $n=4$.

\begin{figure}[H]
\centering
\includegraphics[width=.45\textwidth]{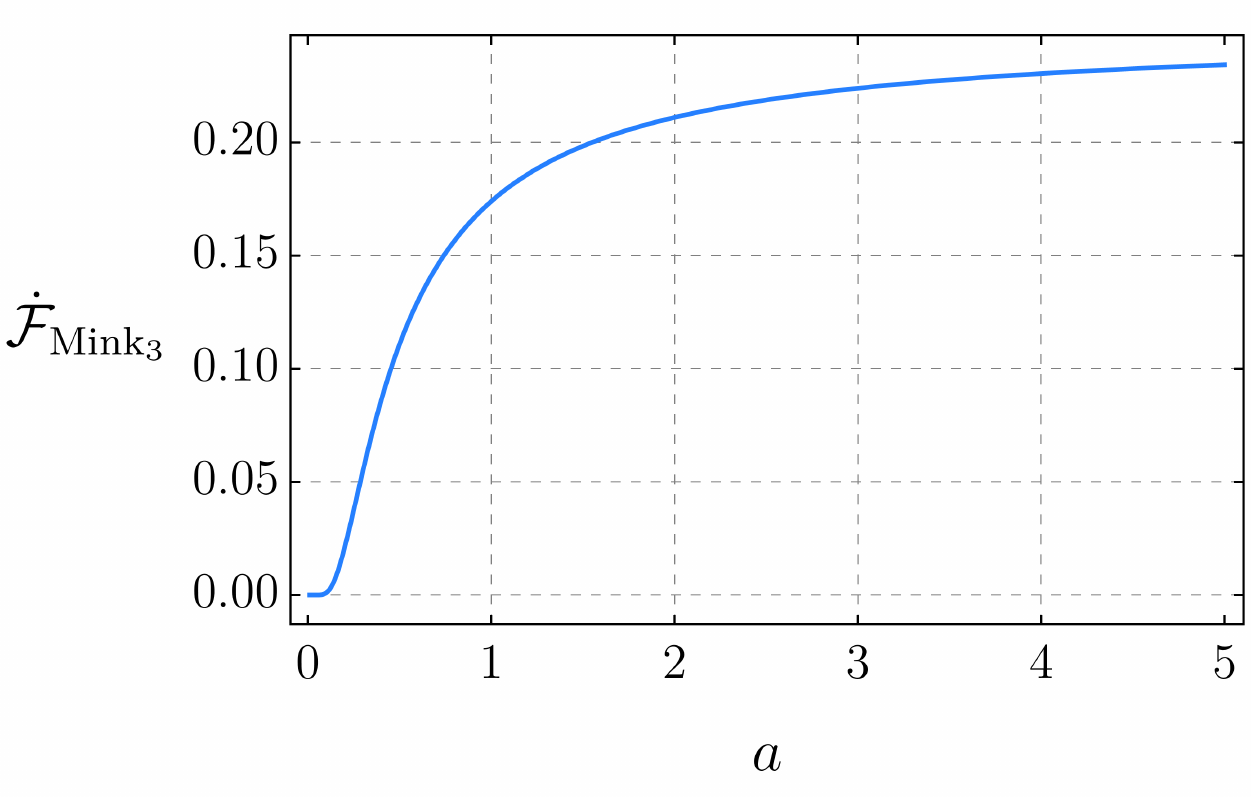}\hspace{.5cm}
	\includegraphics[width=.45\textwidth]{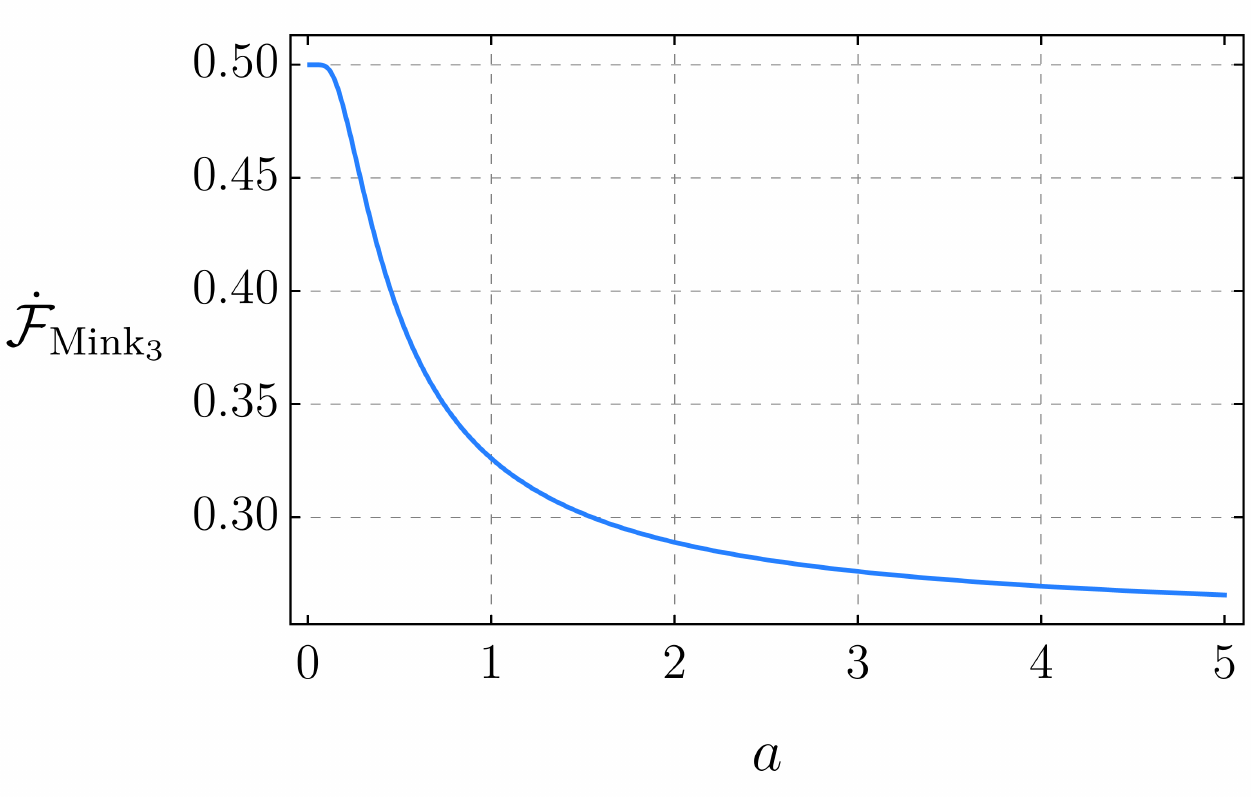}
\caption{The transition rate on the three-dimensional Minkowski spacetime as a function of the proper acceleration of the detector. On the left, for $\Omega=0.1$; on the right, for  for $\Omega=-0.1$.}
\label{fig:transition rate Minkowski 3 as function of a}
\end{figure}

\begin{figure}[H]
\centering
	\includegraphics[width=.45\textwidth]{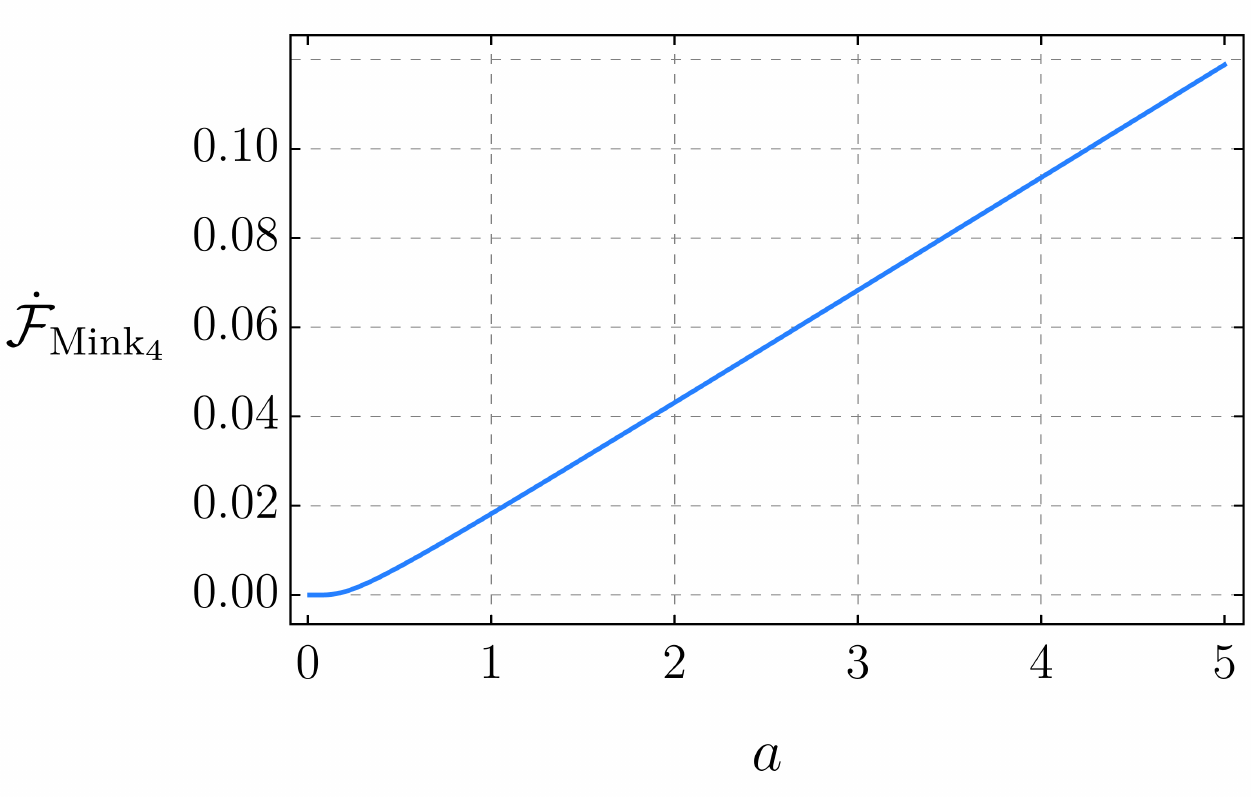}\hspace{.5cm}
	\includegraphics[width=.45\textwidth]{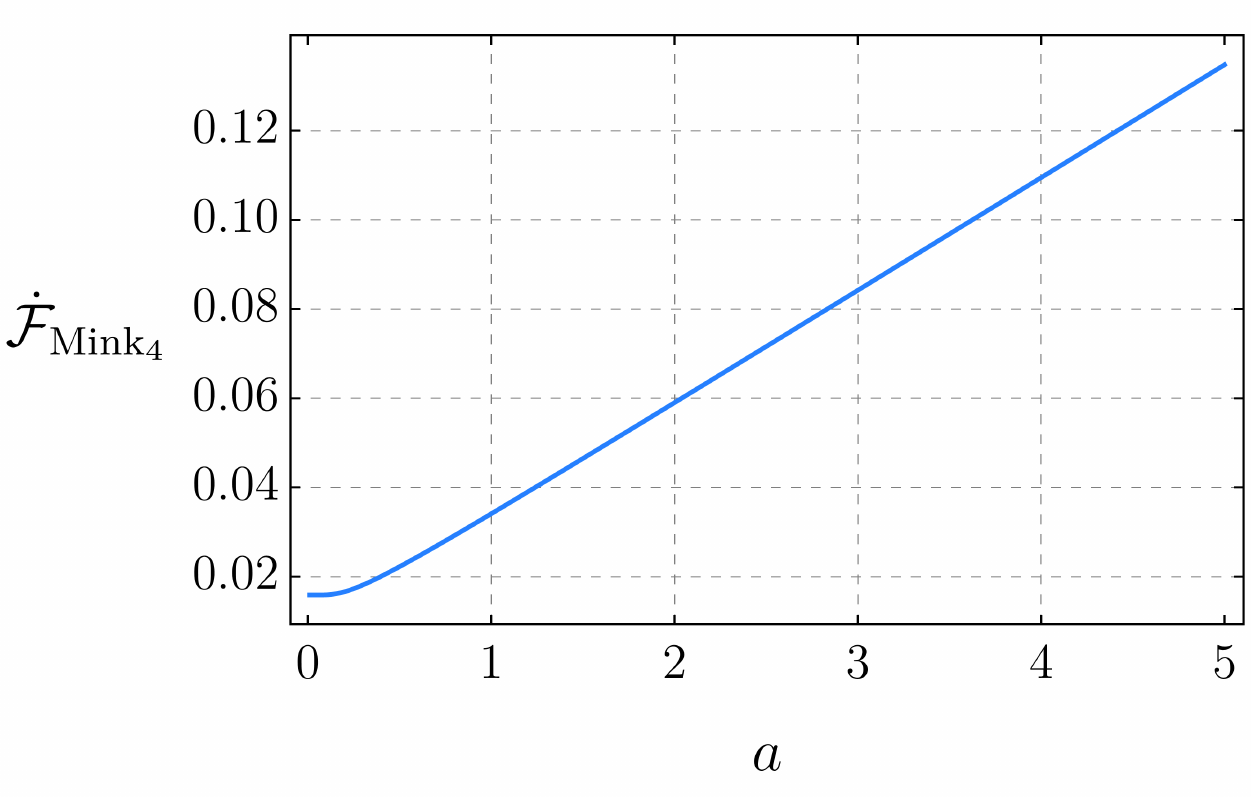}
\caption{ The transition rate on Minkowski spacetime as a function of the proper acceleration of the detector. On the left, for $\Omega=0.1$; on the right, for  for $\Omega=-0.1$.}
\label{fig:transition rate Minkowski 4 5 as function of a}
\end{figure}

With respect to the proper acceleration of the detector, the transition rate is significantly different on the three-dimensional Minkowski spacetime when compared to the higher dimensional counterparts. It is increasing for a positive energy gap, but decreasing for a negative energy gap, as shown in Figure \ref{fig:transition rate Minkowski 3 as function of a}.
On four-dimensional Minkowski spacetime, we observe that the transition rate is increasing for both positive and negative energy gaps. Figure \ref{fig:transition rate Minkowski 4 5 as function of a} illustrates this scenario. Observe that the same qualitative behaviour would have been obtained considering the five- or the six-dimensional Minkowski spacetime.
It is straightforward to verify that all transition rates in Equation \eqref{eq:transition rate Minkowski 3 to 6} satisfy the detailed balance condition. Hence, we can claim that the detector thermalizes with the field at the Unruh temperature $T_U=\frac{a}{2\pi}$
\begin{equation}
	\label{eq:detailed balance on Mink}
	\dot{\mathcal{F}}_{\text{Mink}_i}(-\Omega)  =	e^{\Omega/{T_U}} \dot{\mathcal{F}}_{\text{Mink}_i}(\Omega), \quad i\in\{3,4,5,6\}.
\end{equation}
The relevant information that we can extract from recalling how an Unruh-DeWitt detector responds to Unruh radiation can be summarized as follows. First and foremost, in the three-dimensional Minkowski spacetime, the transition rate for a negative energy gap decreases with the proper acceleration of the detector, {\em i.e. } the anti-Unruh effect, which is just the flat geometry counterpart of the anti-Hawking effect, occurs. Secondly, in the higher dimensional counterparts, the transition rate is always increasing with respect to the proper acceleration of the detector. Last, with respect to the energy gap, the transition rate is decreasing in all dimensions considered.


\subsection{Transition rate on massless hyperbolic black holes}
	\label{sec:Anti-Hawking effect}

 In this section, we summarize the numerical analyses performed for the transition rates \eqref{eq:transition rate for the ground state_original} and \eqref{eq:transition rate for the KMS state} on three- and four-dimensional massless hyperbolic black holes. We focus on a massless real scalar field, conformally coupled to scalar curvature, which corresponds to $\nu=1/2$, and compatible with Robin boundary conditions at conformal infinity for which $\gamma\in[0,\gamma_0)$. More precisely, in all plots we consider fixed boundary conditions $\gamma\in \{0.50 \pi,0.47\pi,0.40\pi,0.25\pi,0\}$, distinguishable by a consistent choice of representation, respectively: blue, purple, pink, orange and green.
 Moreover, the hyperbolic harmonics are written explicitly in Appendix \ref{sec:The Laplace operator on the hyperbolic space}, and are given by Equations \eqref{eq:eigenfunction laplace 3}, for $n=3$, and \eqref{eq:eigenfunction laplace 4} for $n=4$. First, we show the behavior of the transition rates with respect to the energy gap, in Section \ref{sec:Transition rate as a function of the energy gap}. Then, Section \ref{sec:Transition rate as a function of the local Hawking temperature} contains our main results concerning the anti-Hawking effect.

The principle ensuring that the truncation values of $\ell_{\text{max}}$, and of $m_{1,max}$ in the four-dimensional case, are sufficiently large is numerical stability. We outline below the steps followed. Consider the evaluation of the transition rate in the four-dimensional case. Each point in any plot depends on the set of parameters: $$\{\gamma,\nu,\Omega,T_H(\text{or } z_{\mini{D}}),\theta,\varphi_1,\ell_{\text{max}},m_{1,\text{max}},p_{\text{sum}}, p_{\text{int}}\},$$
where $p_{\text{sum}}$ and $p_{\text{int}}$ are the numerical precisions corresponding respectively to the summation over $m_1$ and to the integration over $\ell$. Since the transition rate is given by the product between a small number (the normalization) and a considerably large one (from $R_\gamma$), the precision settings are not trivial. Moreover, since the evaluation can be computationally heavy---one evaluation can take up to one hour, depending on the set of parameters chosen---we studied thoroughly the necessary precision and truncation values at different scales.

Let us consider as an example the plot on the right-hand side of Figure \ref{fig:n = 3 and 4 transition of Egap KMS}. For each fixed set of parameters
\begin{align*}
	\begin{cases}
	&\gamma \in \{0.50 \pi,0.47\pi,0.40\pi,0.25\pi,0\}, \\
	&\nu=1/2,\\
	&z_{\mini{D}}=1/2,\\
	&\theta=1/\pi,\\
	&\varphi=0,\\
	&\Omega\in\{-0.1, -10\},
\end{cases}
\end{align*}
we first stablish the necessary precision of the summation by evaluating the summand of the transition rate at especially high values of $m_1$ and of $\ell$ (order between $100$ and $300$). For example, for $m_1=\ell=100$, one needs $p_{\text{sum}}=150$. Secondly, we determine $p_{\text{int}}$ in an analogous way. After establishing which precision settings are sufficient for each scale of $m_1$ and $\ell$ and for a given set of the fixed parameters, one merely has to check whether the result is stable under increasing truncation parameters. To our goal we deemed sufficient requiring an accuracy of three decimal points. At $\Omega=-10$, for Neumann boundary condition and before normalization, we found $ \dot{\mathcal{F}} = 0.551$
for all values $m_{1,\text{max}}=\ell_{\text{max}} \in \{20,50,100\}$. Each plot required a separate analysis, and they can be found, summarized, in the notebook available at \cite{git_unruh_deWitt_hyperbolic_bh}.


 \subsubsection{Transition rate as a function of the energy gap}
 	\label{sec:Transition rate as a function of the energy gap}

\begin{figure}[H]
 \centering
 \includegraphics[width=0.45\textwidth]{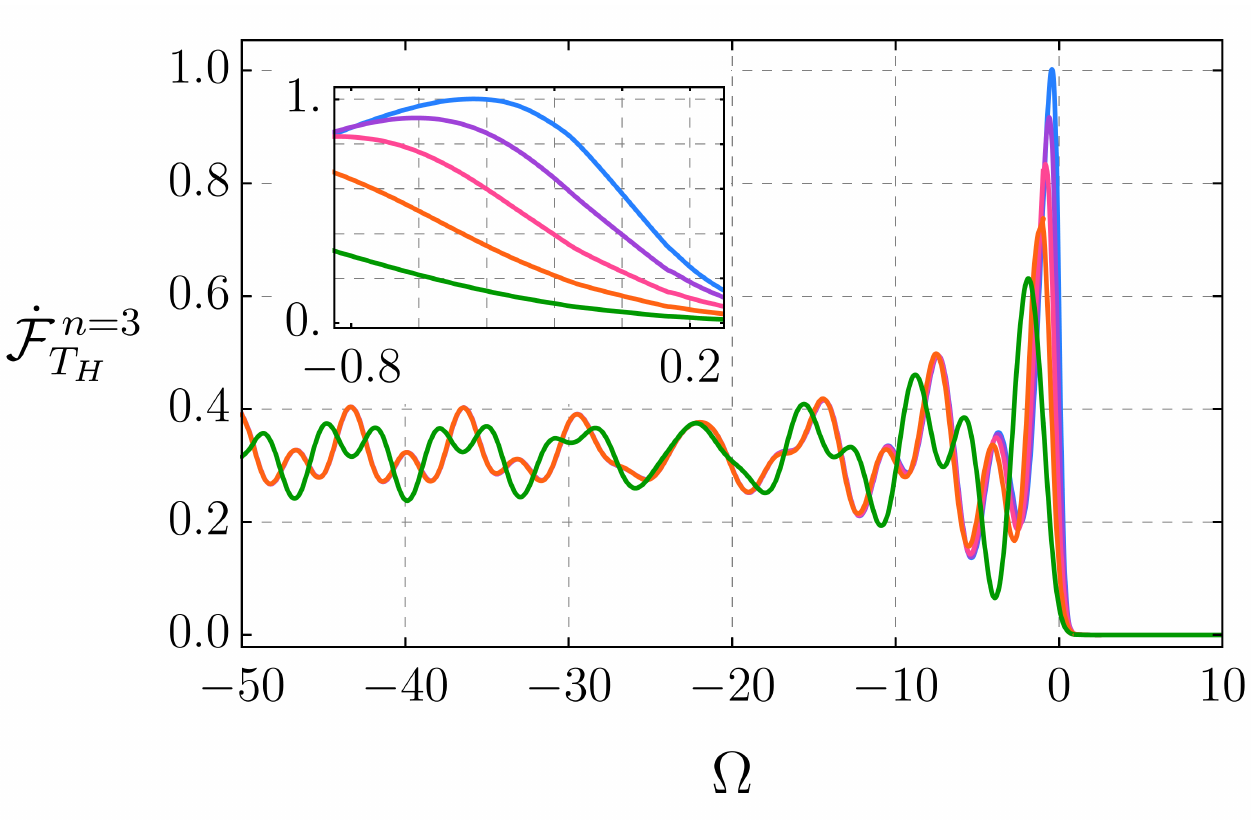}\hspace{.5cm}
 \includegraphics[width=0.45\textwidth]{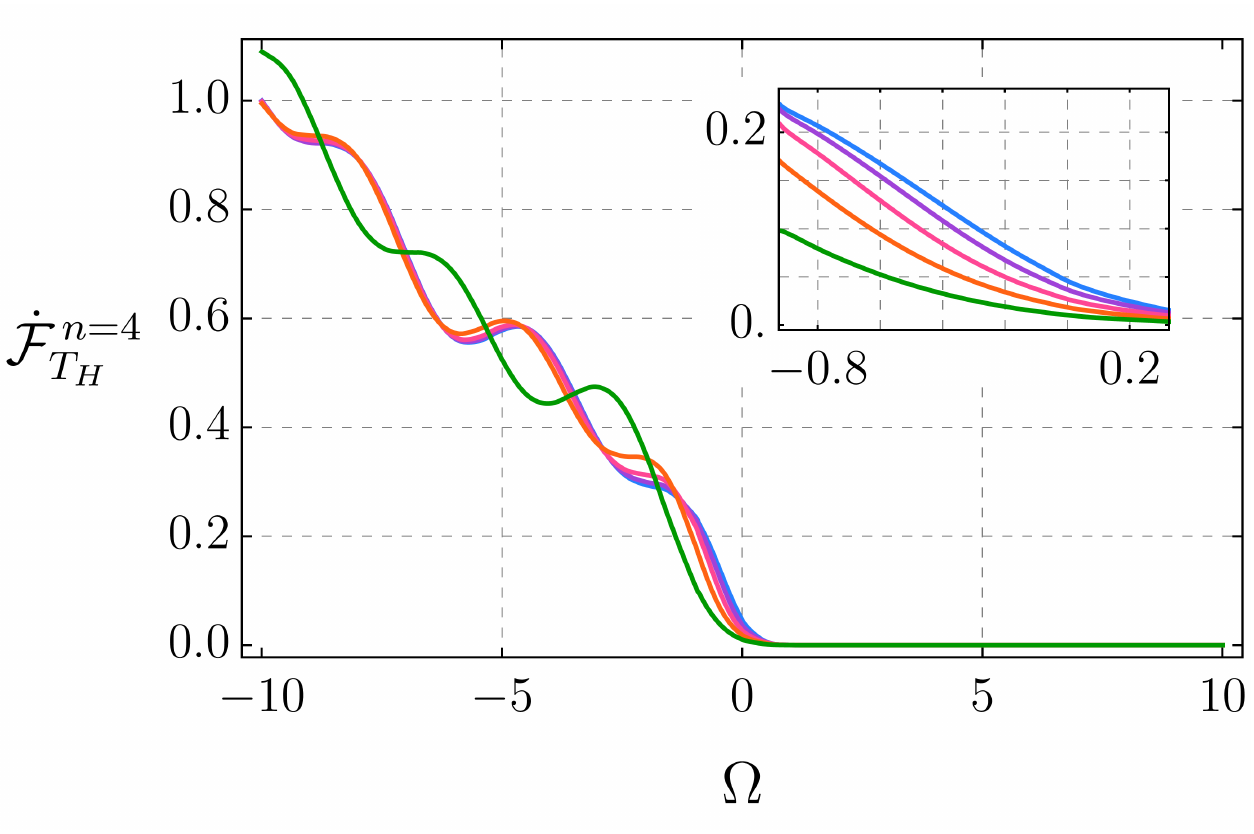}
 \caption{Transition rate as a function of the energy gap for the KMS state at $z_{\mini{D}} = 1/2$, $\theta_{\mini{D}}=\pi^{-1}$ and for different boundary conditions: from top to bottom, $\gamma=(0.50,0.47,0.40,0.25,0)\pi$. On the left, for $n=3$ and with the integration performed up to $\ell=100$; on the right, for $n=4$, at $\varphi_1=0$ and with the summation performed up to $m_1=20$ and integration, up to $\ell=20$.}
 \label{fig:n = 3 and 4 transition of Egap KMS}
 \end{figure}

	With respect to the energy gap, the transition rate for the KMS state behaves on three- and on four-dimensional massless hyperbolic black holes, as shown in Figure \ref{fig:n = 3 and 4 transition of Egap KMS}. On the left, there is the plot for $n=3$ and on the right, that for $n=4$. Observe the similarity with Minkowski spacetime, as shown in Figure \ref{fig:transition rate Minkowski 3 4 5 6 as a function of the energy gap}, but with an extra oscillatory behavior similar to the one observed on a BTZ and on a Schwarzschild black hole, as in \cite{Hodgkinson:2012mr,Hodgkinson:2014iua}.
In the following analysis, we refer to the range $|\Omega|<\pi$ as a small energy gap. In the three-dimensional case, the transition rate displays an oscillatory behaviour for negative energy gaps with absolute values higher than $\sim 5$ and it decreases asymptotically to zero for increasing positive energy gaps. The contrast with the response in the Minkowski counterpart lies, beyond the oscillatory character, in the peak---an apparent global maximum---that it presents for each boundary condition for a small negative energy gap. In the four-dimensional case, the resemblance with the Minkowski counterpart is most evident, also with an extra oscillatory behavior. It is interesting to observe that, in both cases, the transition rate is clearly disparate for different boundary conditions for small energy gaps, but for high energy gaps the disambiguation fades away. For $n=3$, and high energy gaps, we can clearly see that there are basically only two cases: the response for the Dirichlet boundary condition---the green curve---and the response for non-Dirichlet boundary conditions---collapsed in the orange curve.

Numerically, the higher the absolute value of the energy gap, the higher is the cut-off in the $\ell$-integration we have to choose to obtain a stable result. Since, for $n=3$ we considered up to $|\Omega|=50$, we choose $\ell_{\text{max}}=100$. For $n=4$, since we only plotted up to $|\Omega|=10$, we found that $\ell_{\text{max}}=20$ suffices.


 \subsubsection{Transition rate as a function of the local Hawking temperature}
 	\label{sec:Transition rate as a function of the local Hawking temperature}


 In this section, we illustrate the behaviour of the transition rates \eqref{eq:transition rate for the ground state_original} and \eqref{eq:transition rate for the KMS state} with respect to the local Hawking temperature on three- and four-dimensional massless topological black holes still for a real, massless, conformally coupled scalar field.

 As mentioned in the introduction, if the derivative of the transition rate with respect to the local Hawking temperature assumes negative values, which is to say that the transition rate is decreasing with respect to the local Hawking temperature, then we say that the anti-Hawking effect occurs
 As we reviewed in Section \ref{sec:The transition rate on Minkowski spacetime}, the transition rate on Minkowski spacetime is particularly distinct in the three-dimensional case. Recall that the proper acceleration on Rindler trajectories on Minkowski spacetime is proportional to the Unruh temperature, as in Equation \eqref{eq:detailed balance on Mink}.
 As shown in Figure \ref{fig:transition rate Minkowski 3 as function of a}, the transition rate does decrease with increasing proper acceleration when we consider a negative energy gap, which corresponds to considering de-excitation amplitudes. On the other hand, as illustrated in Figure \ref{fig:transition rate Minkowski 4 5 as function of a}, the transition rate, either for excitations or de-excitations, is monotonically increasing with respect to the proper acceleration on Minkowski spacetime of dimensions $4$, $5$ and $6$. What we observe is that on massless hyperbolic black holes, concerning the manifestation of a negative differential effect on the transition rate with respect to the local temperature, it behaves rather similar to how it behaves on Minkowski.

\begin{figure}[H]
 \centering
 \includegraphics[width=0.45\textwidth]{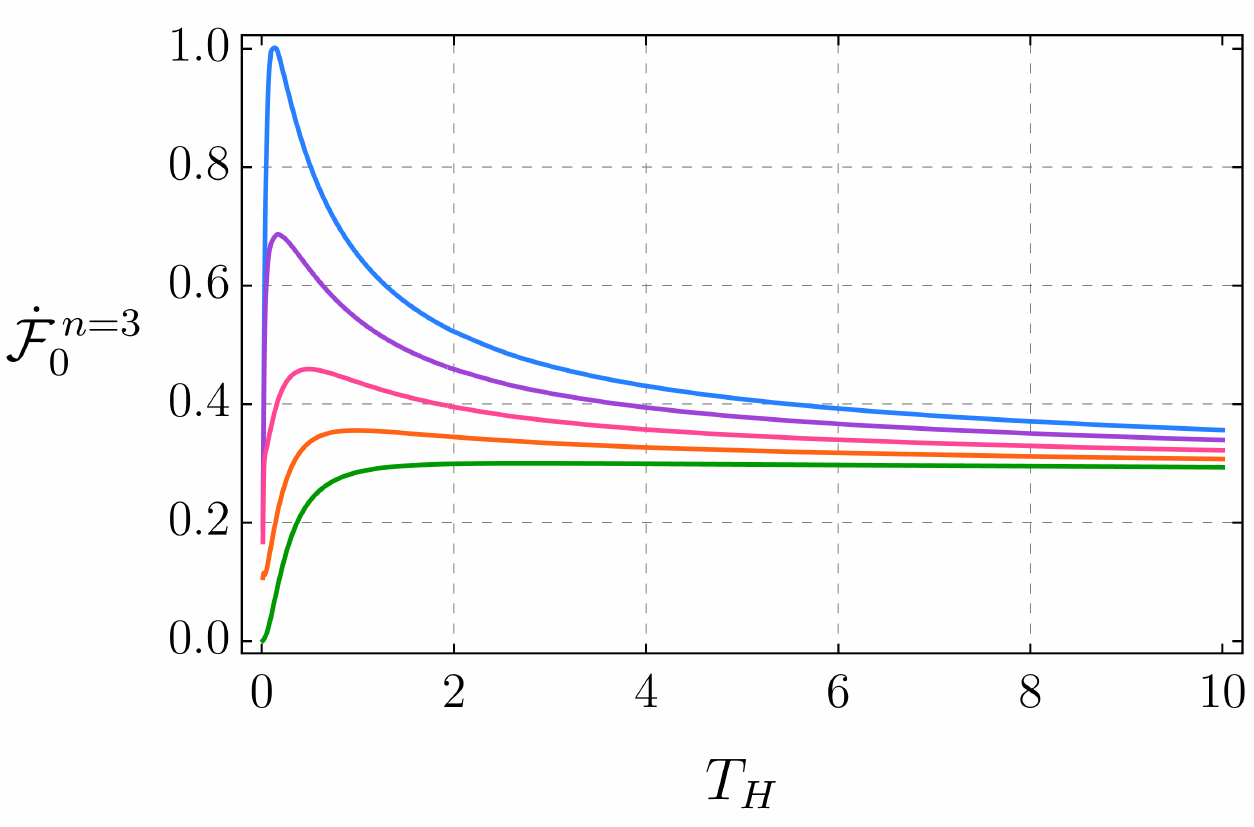}\hspace{.5cm}
 \includegraphics[width=0.45\textwidth]{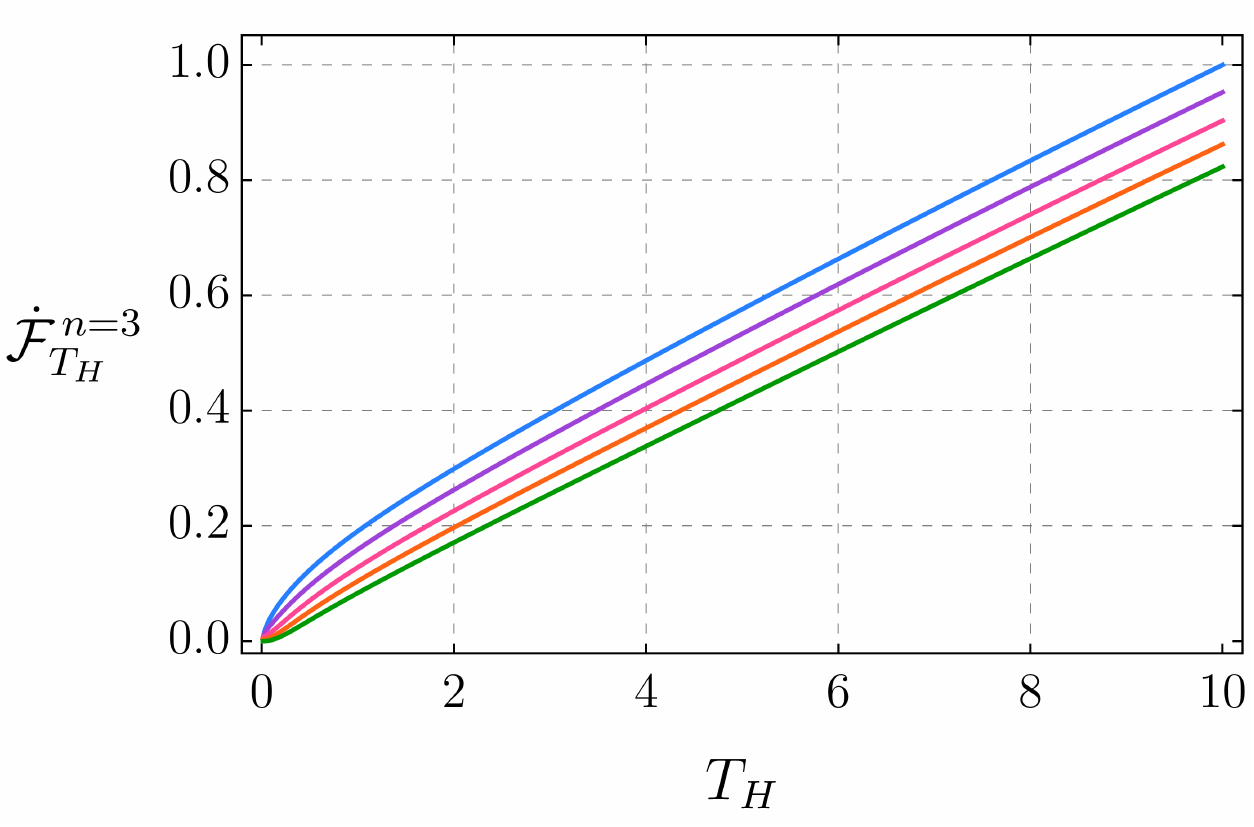}
 \caption{Transition rate, integrated up to $\ell=100$, as a function of the local Hawking temperature on the three-dimensional hyperbolic black hole for $\Omega=-0.1$, $\theta=\pi^{-1}$ and for different boundary conditions. From top to bottom $\gamma=(0.50,0.47,0.40,0.25,0)\pi$. On the left, for the ground state; on the right, for the KMS state.}
 \label{fig:n=3 transition of TH ground and KMS}
 \end{figure}

 As a function of the local Hawking temperature, the transition rate on the three-dimensional massless hyperbolic black hole is given in Figure \ref{fig:n=3 transition of TH ground and KMS}. On the left, we observe that the amplitude of the de-excitations for the ground state does manifest the anti-Hawking effect. It looks very similar to the response on the three-dimensional Minkowski spacetime, as shown in Figure \ref{fig:transition rate Minkowski 3 as function of a}, with a different behavior only for small temperatures $T_H\in(0,1/2)$. The region for which the effect is observed, say $T_H>1/2$, corresponds to $r<1.04944$ and proper accelerations of $a>\sqrt{1+\pi^2}$. 
 For a KMS state however, as on the right of Figure \ref{fig:n=3 transition of TH ground and KMS}, the anti-Hawking effect is not manifest.

\begin{figure}[H]
 \centering
 \includegraphics[width=0.45\textwidth]{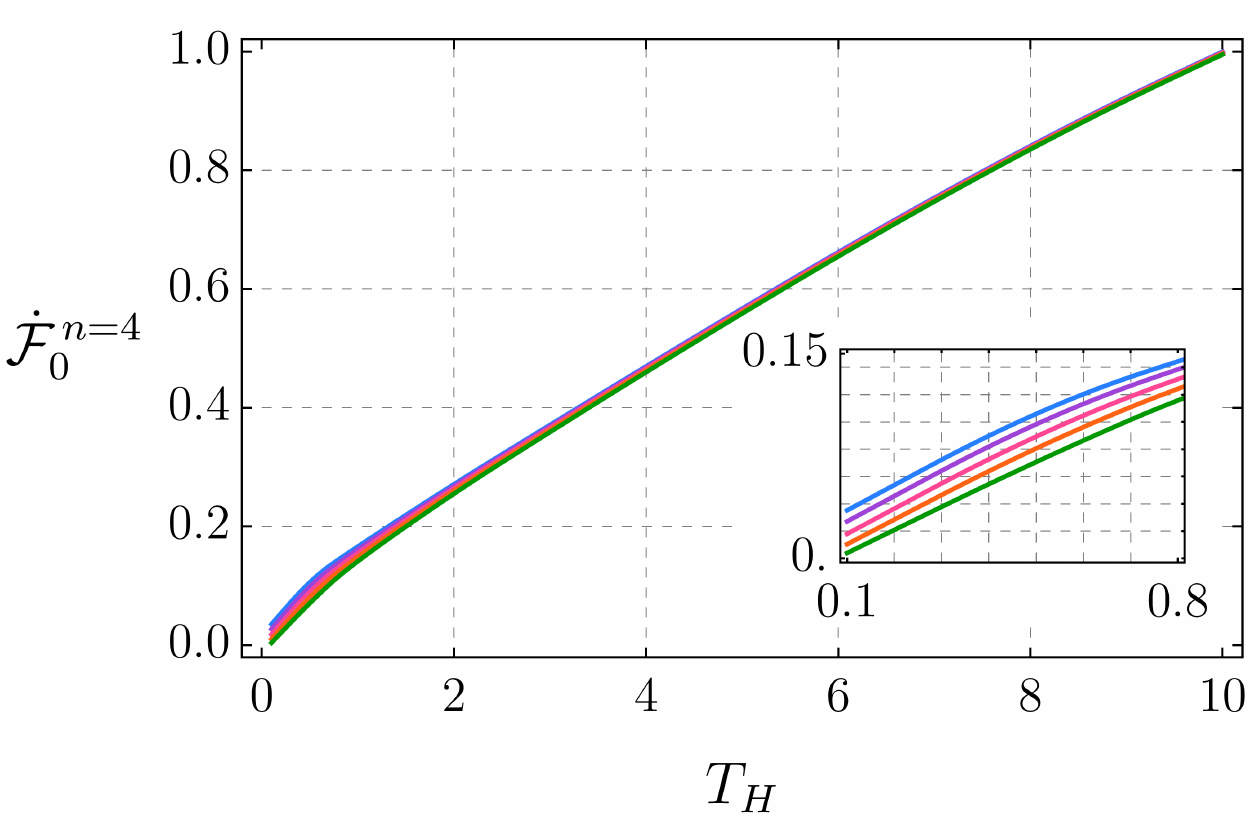}\hspace{.5cm}
 \includegraphics[width=0.45\textwidth]{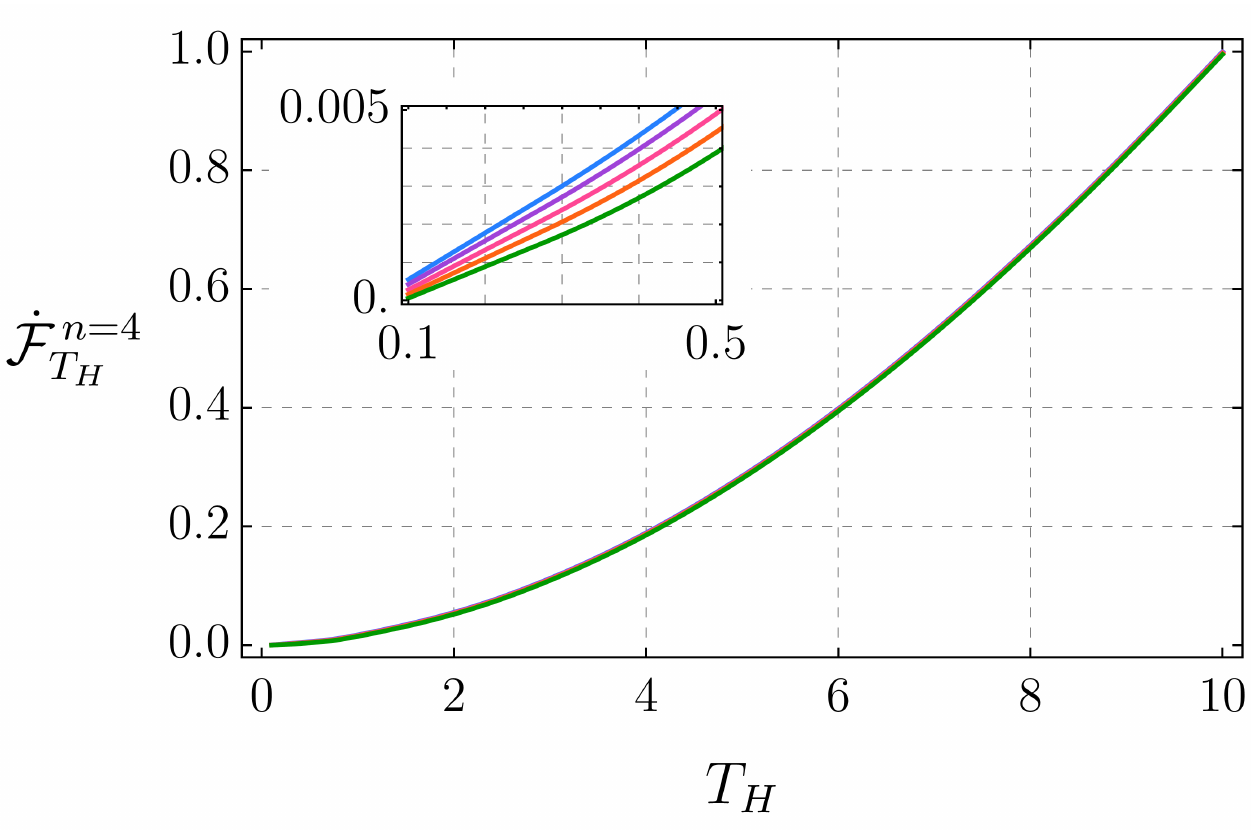}
 \caption{Transition rate as a function of the local Hawking temperature, summed up to $m_1=100$ and integrated up to $\ell=100$, on the four-dimensional hyperbolic black hole for $\Omega=-0.1$, at $\theta_{\mini{D}}=\pi^{-1}$, $\varphi_{1,\mini{D}}=0$ and for different boundary conditions: from top to bottom $\gamma=(0.50,0.47,0.40,0.25,0)\pi$. On the left, for the ground state; on the right, for the KMS state.}
 \label{fig:n=4 transition of TH ground and KMS max 5 and 100}
 \end{figure}

The wave equation does not distinguish a three-dimensional massless hyperbolic black hole from Rindler-AdS$_3$, the universal covering of a static BTZ black hole, also known as Rindler-AdS$_3$ wedge. Indeed, for $n=3$, for positive and negative, constant sectional curvatures, the eigenvalues and eigenfunctions of the Laplacian operator coincide. With that in mind, we remark that the results obtained here are consistent with the ones in \cite{Henderson:2019uqo,Campos:2020twd}. This consistency concerns also the dependence of the anti-Hawking effect from the boundary condition for the ground-state: for $r_h=1$, it is manifest for Neumann boundary condition, but not for the Dirichlet one.
On the four-dimensional massless hyperbolic black hole, the transition rate as a function of the local Hawking temperature is given in Figure \ref{fig:n=4 transition of TH ground and KMS max 5 and 100}.
On the left, we observe the amplitude of the de-excitations for the ground state and we note that the anti-Hawking effect is not manifest. We expect that, for each $T_{H,max}$ chosen, the response will be monotonically increasing with respect to $T_H$ after summing a sufficiently large number of $m_1$ terms, and integrating up to a sufficiently large $\ell$. For $T_{H,max}=10$, $m_{1,max}=100$ and $\ell_{\text{max}}=100$ we have numerical stability and we do not observe any decrease on the transition rate, that is, the anti-Hawking effect does not occur. 
For a KMS state, as we observe on the right column of Figure \ref{fig:n=4 transition of TH ground and KMS max 5 and 100}, the transition rate definitely does not display the anti-Hawking effect. There are two noteworthy features of this plot. One is that it is qualitatively indistinguishable from the one on the four-dimensional Minkowski spacetime, as in Figure \ref{fig:transition rate Minkowski 4 5 as function of a}. The other is that the transition rate barely discriminates the boundary conditions. Yet, as before, we have to keep in mind that $T_H\in(0,1/2)$ corresponds to $r\in(1.04944,\infty)$. The subplots do show that the response functions distinguish the boundary conditions in this interval.
It is also worth mentioning that, focusing on the transition rate for the KMS states, as in Figures \ref{fig:n=3 transition of TH ground and KMS} and \ref{fig:n=4 transition of TH ground and KMS max 5 and 100}, but considering a positive energy gap $\Omega= +0.1$ instead, the behaviour does not change.

\section{Conclusions}
	\label{sec:conclusions}


We have constructed a ground state and a KMS state for a real massive free scalar quantum field theory on $n$-dimensional massless topological black holes. By solving the wave equation, we  obtained an explicit representation for the two-point function, which, by construction, has all relevant physical properties one should require in this context.
The spacetimes considered are not globally hyperbolic, but the prescription here is compatible with Robin boundary conditions at infinity. Moreover, we have written down an explicit expression for the transition rate of an Unruh-DeWitt detector following static trajectories and interacting, for an infinite proper time interval, with the physical states constructed.

As an application of the prescription, we considered the particular case of the massless con\-for\-mal\-ly coupled, real scalar field and we performed a numerical analysis for the three- and four-dimensional cases. We studied the transition rate with respect to the detector's energy gap and obtained, for $n=3$ and $n=4$, a behaviour similar with the one observed on a BTZ black hole and on Schwarzschild black hole, respectively. Since, on Minkowski spacetimes of dimensions $n\in \{3,4,5,6\}$, the anti-Unruh effect is only manifest for spacetime dimension $n=3$, we expect the anti-Hawking effect not to be manifest on massless topological black holes of dimension $n>3$. Indeed, the numerical analysis strongly indicates that the anti-Hawking effect is not manifest on four-dimensional massless topological black holes, for none of the  states considered. Another interesting feature of the transition rate that we noticed when $n\in\{3,4\}$, is that, for higher energy gaps, it only distinguishes between Dirichlet and non-Dirichlet boundary conditions. For $n=4$, we can draw a similar conclusion when analyzing the transition rate as a function of the local Hawking temperature, close to the event horizon.


\section{Acknowledgments}


We are grateful to Jorma Louko for his valuable comments and insights. The work of L.S.C is supported by a PhD scholarship of the University of Pavia, which is gratefully acknowledged. L.S.C. is also grateful for fruitful discussions with Ricardo Correa da Silva. Both authors acknowledge the INFN for travel support, which played a key role in the realization of this project.


\appendix


\section{The Laplace operator on the hyperbolic space}
	\label{sec:The Laplace operator on the hyperbolic space}


	We review the properties of the eigenfunctions and of the eigenvalues of the Laplace operators on the homogeneous hyperbolic manifolds $\Sigma_{n-2}$ introduced in Section \ref{sec: Massless hyperbolic black holes}. Observe that $\Sigma_{n-2}$ is isometric to $H_{(n-2)}^1$, the upper sheet of the two-sheeted hyperboloid---a copy of the $(n-2)$-dimensional hyperbolic space. This corresponds to case F in \cite{Limic}, to which we refer for further details. Note that, for consistency with the rest of this work and clarity, we are using a slightly different notation from \cite{Limic}.

	The hyperbolic space  $H_{n-2}^1$ is diffeomorphic to $\mathbb{R}^{n-2}$, but it has geometrically different properties due to its negative curvature. However, for $n=3$, then $\Sigma_{1}= H_{1}^1$ is not actually a hyperboloid, it is just the open real line parameterized by $\theta\in\mathbb{R}$. In this case, the eigenfunctions of the Laplacian operator are simply plane waves
	\begin{align}
		\label{eq:eigenfunction laplace 3}
		Y^{\ell}(\theta) = &
		\frac{1}{\sqrt{2\pi}}e^{i \ell \theta}, \ell\in\mathbb{R}.
	\end{align}

	Now let us consider the higher dimensional cases. Define  $p\equiv -\frac{i\ell}{2}+\frac{|\ell_{ \{(n-2)/2 \} }|}{2} +\frac{n-3}{4}$ and $q\equiv  |\ell_{ \{ (n-2)/2 \} }|+ \frac{n-2}{2}$ and, for $r\in\{1,2,3,...\}$
	\begin{align}
		\label{eq:brackets notation}
			&\left[\frac{n-2}{2}\right] := \begin{cases}
																				&\frac{n-2}{2}, \text{ if } n -2 = 2r, \\
																						&\frac{n-3}{2}, \text{ if } n -2 = 2r + 1,
																		\end{cases}
			&\left\{\frac{n-2}{2}\right\} := \begin{cases}
																								&\frac{n-2}{2}, \text{ if } n -2 = 2r, \\
																								&\frac{n-1}{2}, \text{ if } n -2 = 2r + 1.
																			\end{cases}
	\end{align}
 For $n>3$, the eigenfunctions of the Laplace operator on $ H_{(n-2)}^1$ are written in terms of Gauss hypergeometric functions and of the $(n-3)$-dimensional spherical harmonics, see \cite[A.1]{Limic}
		\begin{align}
			\label{eq:hyperbolic harmonics n dimensions}
			Y_{m_1,...,m_{ \{(n-2)/2 \} } }^{\ell,\ell_2,...,\ell_{ \{(n-2)/2 \} } }(\bar{\theta}) = & \frac{1}{N^{\frac{1}{2}} } (\tanh \theta)^{ |\ell_{ \{(n-2)/2 \} }| }(\cosh\theta)^{i\ell-\frac{n-3}{2} } \times \nonumber \\
			& \quad\quad\quad \times F(p,p+1/2,q; \tanh^2\theta )  	Y_{m_1,...,m_{ \{ (n-2)/2 \} } }^{\ell_2,...,\ell_{ \{(n-2)/2 \} } }(\bar{\varphi})
		\end{align}
	where $\bar{\varphi}=(\varphi_1,...,\varphi_{n-3})$, $\bar{\theta}=(\theta,\bar{\varphi})$ and
		\begin{align}
			N= \left|  \frac{(2\pi)^{1/2} \Gamma(i\ell)\Gamma\left(q\right) }{\Gamma\left(p+i\ell\right)  \Gamma\left(p+i\ell+1/2\right)  }   \right|^2 .
		\end{align}
	The completeness relations read
		\begin{align}
			\label{eq:CompletenessHyper}
			\int_{ H_{(n-2)}^1} d\mu(\bar{\theta}) \overline{	Y_{m_1,...,m_{ \{(n-2)/2 \} } }^{\ell,\ell_2,...,\ell_{ \{(n-2)/2 \} } }(\bar{\theta})} 	Y_{m'_1,...,m'_{ \{(n-2)/2 \} } }^{\ell',\ell'_2,...,\ell'_{ \{ (n-2)/2 \} } }(\bar{\theta}') =\delta(\ell-\ell')
							\prod\limits_{k=2}^{\{(n-2)/2 \}}\delta_{\ell_k \ell_{k'}}	\prod\limits_{k=1}^{\{(n-2)/2 \}}\delta_{m_k m_{k'}}
		\end{align}
	and
		\begin{align}
			\int_0^\infty d\ell\sum\limits_{S_\ell}  \overline{	Y_{m_1,...,m_{ \{(n-2)/2 \} } }^{\ell,\ell_2,...,\ell_{ \{(n-2)/2 \} } }(\bar{\theta})} 	Y_{m'_1,...,m'_{ \{(n-2)/2 \} } }^{\ell',\ell'_2,...,\ell'_{ \{ (n-2)/2 \} } }(\bar{\theta}')=\delta(\bar{\theta}-\bar{\theta}';\mu).
		\end{align}
	Here $S_\ell$ denotes the collection of all other indices $\ell_2,...,m_{ \{(n-2)/2 \} }$, which are constrained by the relations (A.4) and (A.5) in  \cite{Limic}. The integral kernel of the delta distribution is defined with respect to the measure $\mu$
		\begin{equation}
			\label{eq:measureHypHarmgefegd}
				d\mu(\bar{\theta})=(\cosh\theta)^{n-3} d\theta d\mu(\bar{\varphi}),
		\end{equation}
	for $\theta\in[0,\infty)$, $\varphi_k\in[0,\pi/2),k\in{1,...,r-1}$ and $\varphi_k\in[0,2\pi),k\in{r,...,n-3}$:
		\begin{equation}
			d\mu(\bar{\varphi})=
				\begin{cases}
					\prod\limits_{k=1}^{r-1}\cos \varphi_{k} (\sin\varphi_k)^{2k-1}d\varphi_k \prod\limits_{k=r}^{2r-1} d\varphi_k, &\text{ if }  n-2 = 2r, \nonumber \\
					(\sin\varphi_{r+1})^{2r-1}d\varphi_{r+1} \prod\limits_{k=1}^{r-1}\cos \varphi_k (\sin\varphi_k)^{2k-1}d\varphi_k \prod\limits_{k=r}^{2r} d\varphi_k, &\text{ if }  n-2 = 2r+1. \nonumber
				\end{cases}
		\end{equation}
		With the complex basis above one can construct a real basis, say $\{ Y_1, Y_2\}$, which satisfies analogous completeness relations\\
		\begin{subequations}
			\label{eq:hyperbolic harmonic real basis }
			\begin{equation}
					\label{eq:hyperbolic harmonic real basis RE n dimensions}
						Y_1(\bar{\theta}) = \frac{	Y_{m_1,...,m_{ \{(n-2)/2 \} } }^{\ell,l_2,...,\ell_{ \{(n-2)/2 \} } }(\bar{\theta}) + \overline{	Y_{m_1,...,m_{ \{(n-2)/2 \} } }^{\ell,l_2,...,\ell_{ \{(n-2)/2 \} } }(\bar{\theta})}}{2},
			\end{equation}
			\begin{equation}
				\label{eq:hyperbolic harmonic real basis IM n dimensions}
					Y_2(\bar{\theta}) = \frac{	Y_{m_1,...,m_{ \{(n-2)/2 \} } }^{\ell,l_2,...,\ell_{ \{(n-2)/2 \} } }(\bar{\theta}) - \overline{	Y_{m_1,...,m_{ \{(n-2)/2 \} } }^{\ell,l_2,...,\ell_{ \{(n-2)/2 \} } }(\bar{\theta})}}{2i}.
			\end{equation}
		\end{subequations}

			\noindent In four dimensions, $\bar{\theta}=(\theta,\varphi_1)$, $p=-i\ell/2 + |\ell_1|/2  + 1/4$, $q=|\ell_1|+1$, $\ell_1\equiv m_1$ and \eqref{eq:hyperbolic harmonics n dimensions} reads
			\begin{align}
				\label{eq:eigenfunction laplace 4}
				Y^{\ell}_{m_1}(\theta,\varphi_1) = &
				 \frac{1}{N^{\frac{1}{2}} } (\tanh \theta)^{ |\ell_1| }(\cosh\theta)^{i\ell-\frac{1}{2} } F(p,p+1/2,q; \tanh^2\theta )  	Y_{m_1}(\varphi_1),
			\end{align}
			where $Y_{m_1}(\varphi_1)=e^{i m_1 \varphi_1}$. Since these eigenfunctions depend also on $m_1$, we have that, beyond the integral in $\ell$, the expressions for the two-point functions and for the transition rates \eqref{eq:transition rate for the ground state_original} and \eqref{eq:transition rate for the KMS state} also have an implicit sum over $m_1\in\mathbb{Z}$.



\section{On the finiteness of the transition rate}
	\label{sec:The transition rate is a well-defined positive quantity}


\subsection{Asymptotic expansion of the hypergeometric functions}
\label{sec:The hypergeometric solutions diverge exponentially}

The hypergeometric functions that appear in Equation (\ref{eq:transition rate for the KMS state}) diverge exponentially as $\ell\rightarrow\infty$. The $\ell$-dependence of the solutions $R_{1(1)}$ and $R_{2(1)}$ appears in the parameters of the hypergeometric solutions. Now, the standard large-parameter asymptotic expansion one can find in \cite[(15.12.5)]{NIST}, which comes from \cite{jones2001} and is written in terms of $(1-z')/2$ for $z'=2z-1$, does not hold for our case: $|\text{ph} (z'-1)|=\pi$, where $\text{ph}$ stands for the phase of a complex number. Yet an analytic extension which does include our case was studied in \cite{farid2014uniform}. Let $\sigma  := \eta/4 + i \sqrt{|g_{00}|}\Omega/2 $, $\xi=\ln(-z'-i\sqrt{1-z'^2})$ and let $z'=2z-1$. The asymptotic expression \cite[(3.2)]{farid2014uniform}, for large $\ell$, gives
	\begin{align}
		F\left(\sigma + i\ell/2 ,\sigma - i\ell/2; \eta/2; 1-z\right)\sim C(\sigma,\eta,z') \frac{e^{ \ell/2 (\pi - i\xi) } }{\sqrt{ i \ell/2}} \frac{\Gamma(i\ell/2+1+\sigma-\eta/2)}{\Gamma(i\ell/2+\sigma)}
	\end{align}
where the $\ell$-independent coefficient is given by $$ C(\sigma, \eta, z') := \frac{e^{ i \pi(\eta-1)/4 } \Gamma(\eta/2)}{2^{\sigma-1}\sqrt{\pi}}\frac{(1+z')^{\eta/4-\sigma-1/4}}{(1-z')^{\eta/4-1/4}}.$$
In the massless conformally coupled case, that is $\nu=1/2$ and for all $n\geq3$, we have that for $\eta=3$, the expansion above applies to the solution in Equation (\ref{eq:sol11hypergeo}) and for $\eta=1$, to that in Equation (\ref{eq:sol21hypergeo}). Yet, the infinite $\ell$-limit does not depend on $\eta$. Using the asymptotic expansion of the Gamma function \eqref{eq:asympGamma}, consequence of Stirling’s formula,
 \begin{equation}
 	\label{eq:asympGamma}
	 \frac{\Gamma(z+p)}{\Gamma(z+q)} \overset{|z|\rightarrow\infty}{\sim}z^{p-q} \quad\text{ for }p,q,z\in\mathbb{C} \text{ and }|\text{ph} \, z|<\pi,
 \end{equation}
and noticing that $|-z'-i\sqrt{1-z'^2}|=1$ we have $\xi= i \varphi$ for $\varphi = \text{ph}(-z'-i\sqrt{1-z'^2})\in(-\pi,0)$, then
	\begin{align}
		F\left(\sigma + i\ell/2 ,\sigma - i\ell/2; \eta/2; 1-z\right)\sim D(\sigma,\eta,z')e^{\ell (\pi + \varphi)/2}  \ell^{(1-\eta)/2}
	\end{align}
for $D(\sigma,\eta,z'):= (i/2)^{(1-\eta)/2} C(\sigma,\eta,z') $. Therefore, both hypergeometric solutions grow exponentially with $e^{\frac{\pi}{2}\ell }$ as $\ell\rightarrow \infty$, and so does $R_\gamma$.


\subsection{Asymptotic expansion of the normalization factor}
\label{sec:Asymptotic expansion of the normalization factor}

\begin{figure}[H]
		 \centering
				\hspace{-.5cm}\includegraphics[width=.45\textwidth]{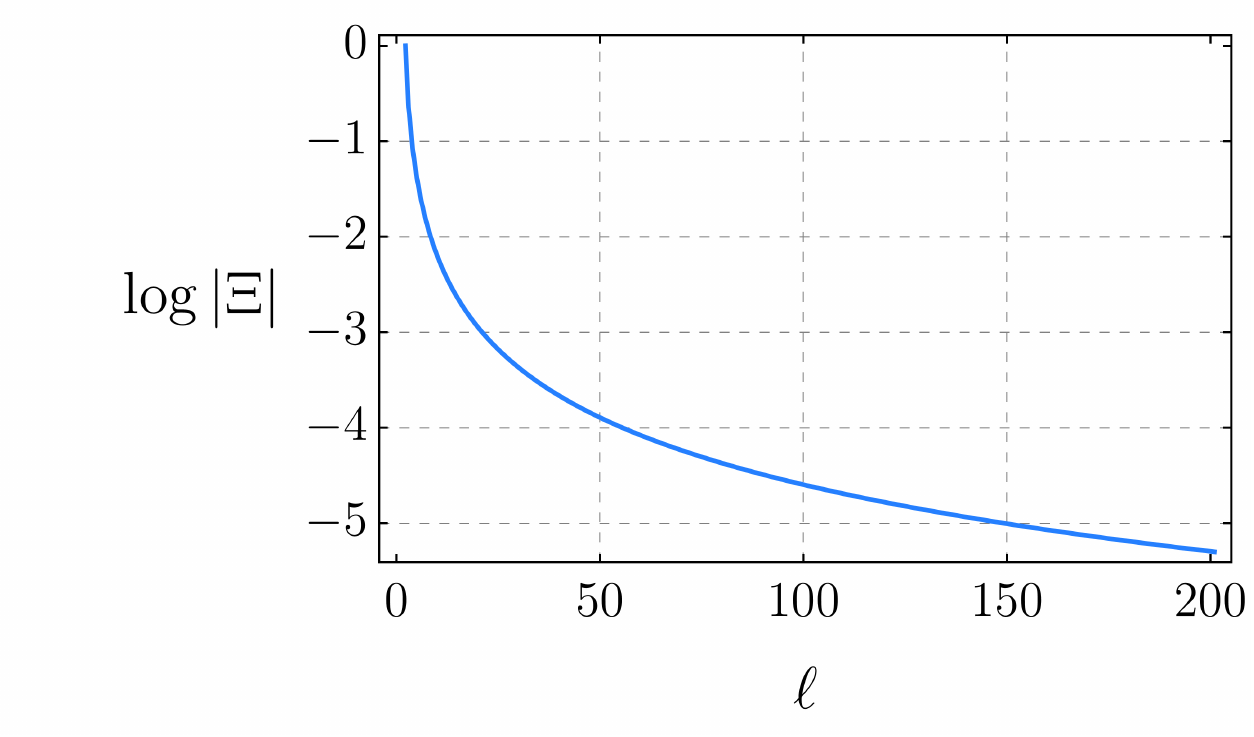}\hspace{.5cm}
				\includegraphics[width=.45\textwidth]{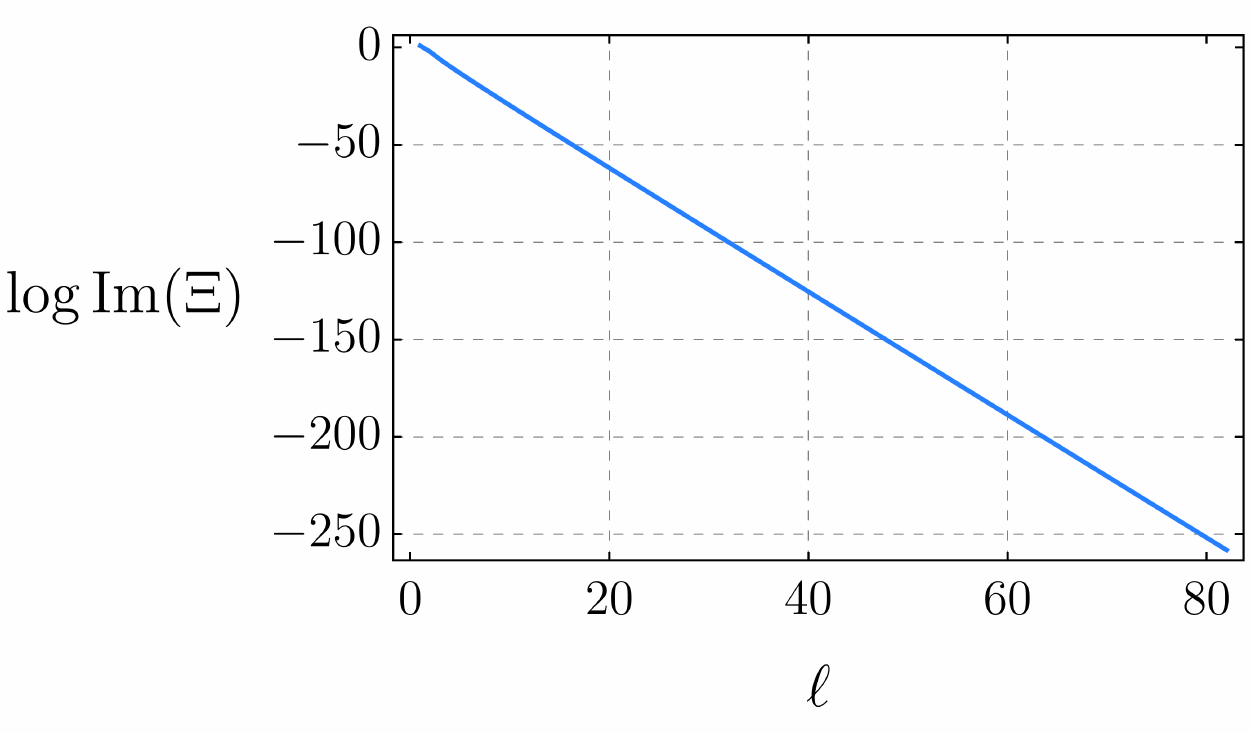}
				\caption{On the left, absolute value of $\Xi$ in log scale as a function of $\ell$ for $\Omega=0.1$ at $z_{\mini{D}}=0.9$ and with $\gamma=\pi/4$. On the right, the corresponding imaginary part. The absolute value goes to zero linearly, but, seemingly, the imaginary part goes to zero exponentially. Recall that $\Xi$ does not depend on $n$.}
		 \label{fig:itoscillates}
		\end{figure}

Consider the normalization $\mathcal{N}^{\mini{-1}}$ as per Equation \eqref{eq:normwithXi}. For convenience, consider $\nu=1/2$ and define $s_1 := \frac{ (\ell +  \sqrt{|g_{00}|}\Omega)}{2}$ and $s_2 :=  \frac{ (\ell -  \sqrt{|g_{00}|}\Omega)}{2}$. The dependence of $\mathcal{N}^{\mini{-1}}$ on $\ell$ comes from the function $\Xi\equiv\Xi(\omega)=\frac{B}{A}$, {\it cf.} Equation \eqref{eq:CoefA} and \eqref{eq:CoefB}
	\begin{align}
	 \Xi
			 &= -\frac{1}{2} \frac{\Gamma( 1/4 -i s_2)}{\Gamma(3/4 - i s_2)}\frac{\Gamma( 1/4 + i s_1)}{\Gamma( 3/4 + i s_1)}.
 \end{align}
The imaginary part of $\Xi$ can be written as
\begin{align}
	\Imag(\Xi)
 		 &= -\frac{1}{4 i}  \Bigg\{ \left[  \frac{\Gamma( 1/4 -i s_2)}{\Gamma(3/4 - i s_2)}\frac{\Gamma( 1/4 + i s_1)}{\Gamma( 3/4 + i s_1)}   \right]  - \left[  \frac{\Gamma( 1/4 -i s_2)}{\Gamma(3/4 - i s_2)}\frac{\Gamma( 1/4 + i s_1)}{\Gamma( 3/4 + i s_1)}   \right]^\ast \Bigg\} \nonumber \\
		 &=-\frac{1}{4 i}\Bigg\{
-\frac{8 i \pi ^2 \sinh (\pi  \sqrt{|g_{00}|}\Omega)}{\cosh (2 \pi  \ell)+\cosh (2 \pi  \sqrt{|g_{00}|}\Omega)} E(\ell)
		 \Bigg\} \nonumber \\
		 &= \frac{ c_1}{(\cosh (2 \pi  \ell)+c_2)} E(\ell) \nonumber \\
		 & \overset{\ell\rightarrow\infty}{\sim }e^{-2\pi\ell} E(\ell),
\end{align}
where the constants are $c_1 := 2\pi^2 \sinh (\pi  \sqrt{|g_{00}|}\Omega)$, $c_2 := \cosh (2 \pi  \sqrt{|g_{00}|}\Omega)$, while $E(\ell)$ is defined by
\begin{equation}
E(\ell):=\frac{1}{ \Gamma( 3/4 + i s_1 )\Gamma( 3/4 - i s_1 )\Gamma(3/4 + i s_2) \Gamma(3/4 - i s_2 )}.
\end{equation}
Using that \cite[Eq.(5.11.9)]{NIST}
\begin{equation}
 |\Gamma(x+iy)|\overset{y\rightarrow \pm \infty} \sim |y|^{x-1/2}e^{-\pi |y|/2 },
\end{equation}
we have that the asymptotic behaviour of $E(\ell)$ is
\begin{equation}
	\label{eq:asymptotic of E(l)}
E(\ell)\overset{y\rightarrow \pm \infty} \sim (|s_1|^{-1/4} e^{\frac{\pi}{2}|s_1|})(|-s_1|^{-1/4} e^{\frac{\pi}{2}|-s_1|}) (|s_2|^{-1/4} e^{\frac{\pi}{2}|s_2|})(|-s_2|^{-1/4} e^{\frac{\pi}{2}|-s_2|})  \sim  \frac{e^{\pi \ell}}{\ell}.
\end{equation}
Therefore, the imaginary part of $\Xi$ decreases exponentially with $\pi\ell$. Analogously, we have for the real part
\begin{align}
	\Real(\Xi)
 		 &= -\frac{1}{4 }  \Bigg\{ \left[ \frac{\Gamma( 1/4 -i s_2)}{\Gamma(3/4 - i s_2)}\frac{\Gamma( 1/4 + i s_1)}{\Gamma( 3/4 + i s_1)}\right]  + \left[ \frac{\Gamma( 1/4 -i s_2)}{\Gamma(3/4 - i s_2)}\frac{\Gamma( 1/4 + i s_1)}{\Gamma( 3/4 + i s_1)}\right]^\ast \Bigg\} \nonumber \\
		 &=-\frac{1}{4}\Bigg\{
\frac{8  \pi ^2 \cosh (\pi \ell )}{\cosh (2 \pi  \ell)+\cosh (2 \pi  \sqrt{|g_{00}|}\Omega)} E(\ell)
		 \Bigg\} \nonumber \\
		 &=-2\pi^2 \frac{ \cosh (\pi \ell )}{\cosh (2 \pi  \ell)+c_2} E(\ell) \nonumber \\
		 & \overset{\ell\rightarrow\infty}{\sim } e^{-\pi\ell} E(\ell).
\end{align}
On account of Equation \eqref{eq:asymptotic of E(l)}, we obtain that the real part of $\Xi$, and thus its absolute value as well, decreases linearly with $\ell$. This behaviour for $\Xi$ is indeed observed numerically. We conclude that $\mathcal{N}^{\mini{-1}}$ also goes exponentially to zero. Plot \ref{fig:itoscillates} bellow illustrates the behavior of $|\Xi|$ and of its imaginary part.
Note that the rate of exponential decay of the imaginary part of $\Xi$, ruled by $\pi$, is, indeed, considerably larger than that of the exponential growth of $R_\gamma$ as a function of $\ell$, ruled by $\pi/2$.

\subsection{Vanishing imaginary part for $R_\gamma$}
\label{sec:Vanishing imaginart part for Rgamma}


Consider the solution $R_{1(1)}(z)=z^{\alpha_+}(1-z)^{\beta_+} h_{1(1)}(z)$, where $h_{1(1)}(z)$ is given implicitly by \eqref{eq:sol11hypergeo} for $a,b,c$ as in \eqref{eq:parameters a b and c of hypergeometric equation} and $\alpha_+$ as in \eqref{eq:auxiliary parameters alpha and beta plus and minus}. Assuming $\omega^2\in\mathbb{R}$, its complex conjugate is given by
\begin{equation}
\overline{R_{1(1)}(z)}= z^{-\alpha_+}(1-z)^{\beta_+} \overline{h_{1(1)}(z)}.
\end{equation}
Using identity \cite[eq.(15.10.13)]{NIST},
\begin{equation}
F\left(a,b,a+b+1-c;1-z\right) = z^{1-c}F(a-c+1,b+1-c;a+b-c+1;1-z),
\end{equation}
and noticing that $1-c=-2\alpha_+$, $a-c+1=\overline{b}$, $b-c+1=\overline{a}$ and $a+b-c=\nu$, we get
\begin{align}
\overline{h_{1(1)}(z)} &= \left[z^{1-c}F(a-c+1,b-c+1;a+b-c+1;1-z)\right]^\ast \nonumber\\
										&= z^{2\alpha_+}F(b,a; 1+\nu ;1-z).
\end{align}
Therefore, it holds
\begin{equation}
\overline{R_{1(1)}(z)}= z^{\alpha_+}(1-z)^{\beta_+}h_{1(1)}(z)=R_{1(1)}(z)\in\mathbb{R}.
\end{equation}

An analogous argument holds for $R_{2(1)}(z)$. Using the second equality within  \cite[eq.(15.10.14)]{NIST}, and since $1-\bar{a}=c-b$ and $1-\bar{b}=c-a$, we have
\begin{align}
\overline{h_{2(1)}(z)} &= \left[z^{-\nu}F(c-a,c-b;1-\nu;1-z)\right]^\ast \nonumber\\
										 &=     z^{-\nu}z^{2\alpha_+}   F( c-a , c-b ; 1-\nu;1-z).
\end{align}
Which implies
\begin{align}
\overline{R_{2(1)}(z)}&=z^{\alpha_+}(1-z)^{\beta_+} f_{2(1)}(z)=R_{2(1)}(z)\in\mathbb{R}.
\end{align}
Therefore, $R_\gamma(z)$, which is a linear combination of $R_{1(1)}(z)$ and $R_{2(1)}(z)$ with real-valued coefficients given by Equation \eqref{eq:Robin_solution}, is also real-valued.


\subsection{On the convergence of the integral in $\ell$}


The transition rates, given by Equations \eqref{eq:transition rate for the ground state_original} and \eqref{eq:transition rate for the KMS state}, are written as integrals in $\ell$.
In Section \ref{sec:Asymptotic expansion of the normalization factor}, we showed that, for large $\ell$, the integrand decays exponentially with $\ell$ despite the fact that the hypergeometric solutions diverge exponentially, which was shown in Section \ref{sec:The hypergeometric solutions diverge exponentially}. Here, we give a simple argument proving that the integral in $\ell$ is convergent.

Suppose $\zeta(\ell)$ is a function such that there exists a large $\ell$ value, say $\Lambda$, such that $\zeta(\ell) \overset{\ell \geq \Lambda}{\sim} e^{-\ell}$. Therefore
			\begin{align}
				\left|\int_0^\infty d\ell \zeta(\ell) \right|
			    \leq\int_{0}^{\Lambda}d\ell	\left|\zeta(\ell) \right| + \int_{\Lambda}^{+\infty}d\ell \left|\zeta(\ell) \right| 
				 = \int_{0}^{\Lambda}d\ell 	\left|\zeta(\ell) \right| - e^{-\Lambda} 
			   <\infty.
		\end{align}

Since the argument above applies to the transition rates, then the integrals in $\ell$ converge. Since the normalization in Equation \eqref{eq:normwithXi} and the solutions $R_\gamma(z)$ are real-valued, as shown in Section \ref{sec:Vanishing imaginart part for Rgamma}, we conclude that the transition rate as in Equation \eqref{eq:transition rate for the ground state_original} and \eqref{eq:transition rate for the KMS state} are real-valued, finite numbers.



\end{document}